\documentclass[a4paper,fleqn]{cas-sc}

\usepackage[authoryear]{natbib}
\usepackage{graphicx} 
\usepackage{float}
\usepackage{algorithm}  
\usepackage{algpseudocode}
\usepackage{color}
\usepackage{setspace}
\usepackage{hyperref}
\def\tsc#1{\csdef{#1}{\textsc{\lowercase{#1}}\xspace}}
\tsc{WGM}
\tsc{QE}
\tsc{EP}
\tsc{PMS}
\tsc{BEC}
\tsc{DE}
\usepackage{lineno}
\begin{document}
\let\WriteBookmarks\relax
\def\floatpagepagefraction{1}
\def\textpagefraction{.001}
\let\printorcid\relax
\shorttitle{Statistically controllable microstructure reconstruction framework}
\shortauthors{Zhenchuan Ma, Qizhi Teng, Pengcheng Yan, et al}

\title [mode = title]{Statistically controllable microstructure reconstruction framework for heterogeneous materials using sliced-Wasserstein metric and neural networks}

\author[a]{Zhenchuan Ma}[style=chinese]
%\ead{mazhenchuan@stu.scu.edu.cn}
\credit{Data curation, Methodology, Software, Writing – original draft}

\author[a]{Qizhi Teng}[style=chinese]
\credit{Supervision, Funding acquisition}

\author[a]{Pengcheng Yan}[style=chinese]
\credit{Investigation, Methodology, Validation}

\author[a]{Lindong Li}[style=chinese]
\credit{Investigation, Methodology, Software}

\address[a]{College of Electronics and Information Engineering, Sichuan University, China}

\author[b]{Kirill M. Gerke}[style=chinese]
\credit{Investigation, Methodology}

\author[b]{Marina V. Karsanina}[style=chinese]
\credit{Methodology}

\address[b]{Moscow Institute of Physics and Technology, Moscow, Russia}

\author[a]{Xiaohai He}[style=chinese]
\cormark[1]
\credit{Methodology, Supervision, Funding acquisition}
%\fnmark[]
%\cortext[1]{Corresponding author: Xiaohai He. E-mail: hxh@scu.edu.cn}
%\fntext{Corresponding author: Xiaohai He. E-mail: hxh@scu.edu.cn}

\begin{abstract}
Heterogeneous porous materials play a crucial role in various engineering systems. Microstructure characterization and reconstruction provide effective means for modeling these materials, which are critical for conducting physical property simulations, structure–property linkage studies, and enhancing their performance across different applications. To achieve superior controllability and applicability with small sample sizes, we propose a statistically controllable microstructure reconstruction framework that integrates neural networks with sliced-Wasserstein metric. Specifically, our approach leverages local pattern distribution for microstructure characterization and employs a controlled sampling strategy to generate target distributions that satisfy given conditional parameters. A neural network-based model establishes the mapping from the input distribution to the target local pattern distribution, enabling microstructure reconstruction. Combinations of sliced-Wasserstein metric and gradient optimization techniques minimize the distance between these distributions, leading to a stable and reliable model. Our method can perform stochastic and controllable reconstruction tasks even with small sample sizes. Additionally, it can generate large-size (e.g. 512 and 1024) 3D microstructures using a chunking strategy. By introducing spatial location masks, our method excels at generating spatially heterogeneous and complex microstructures. We conducted experiments on stochastic reconstruction, controllable reconstruction, heterogeneous reconstruction, and large-size microstructure reconstruction across various materials. Comparative analysis through visualization, statistical measures, and physical property simulations demonstrates the effectiveness, providing new insights and possibilities for research on structure–property linkage and material inverse design.
\end{abstract}
 
%\begin{highlights}
%\item A statistically controllable framework for small sample data.
%\item Controlled strategy for generating microstructures with specified phase volume fractions.
%\item Local–global decoupling reconstruction for complex, heterogeneous microstructures.
%\item Capable of generating large-size (e.g., 1024) 3D microstructures.
%\end{highlights}


\maketitle 

\printcredits

\doublespacing

\section{Introduction}\label{sec1}

%\doublespacing
Heterogeneous porous materials, such as rock, silica, concrete, superalloys, and synthetic ceramics, are characterized by a solid matrix interspersed with pore spaces or other distinct phases. These materials play a critical role across various engineering applications, including oil and gas extraction from reservoirs \citep{1_farajzadeh2012foam}, geological sequestration of carbon dioxide \citep{2_shukla2010review}, filtration through porous membranes \citep{3_hotza2020silicon}, heat exchange enhancement via metal foams \citep{4_zhang2017melting}, separation processes within lithium-ion batteries \citep{5_lagadec2019characterization}, and noise reduction and heat insulation panels in building components \citep{CBM-hemmati2024acoustic}. Establishing the relationships between structure and properties in these materials is crucial for understanding and controlling transport phenomena and reaction mechanisms, which in turn enhances their performance across diverse systems \citep{6_chen2022pore}.

As the foundation for subsequent research, both direct imaging techniques and microstructure characterization and reconstruction methods are essential for establishing digital models of the microstructure for heterogeneous porous materials. Three-dimensional (3D) direct imaging techniques \citep{8_kizilyaprak2014focused,7_withers2021x} enable detailed visualization of these materials and the creation of realistic and reliable digital microstructure models. However, the high costs and operational complexity associated with these techniques limit their widespread application. In contrast, microstructure characterization and reconstruction methods \citep{9_bostanabad2018computational} offer a rapid, cost-effective, and efficient alternative for modeling microstructures. These methods can generate 3D structures with similar statistical or mechanistic properties based on two-dimensional (2D) images. Importantly, they possess the unique capability to produce specific microstructures tailored to given characteristic and property parameters. This feature facilitates the generation of widely available microstructure data, which is crucial for studying structure–property linkages \citep{11_shang2023tailoring} and supporting material inverse design \citep{11_1wang2024diffmat}.

\begin{figure}%[htbp]
	\centering
        \includegraphics[scale=.32]{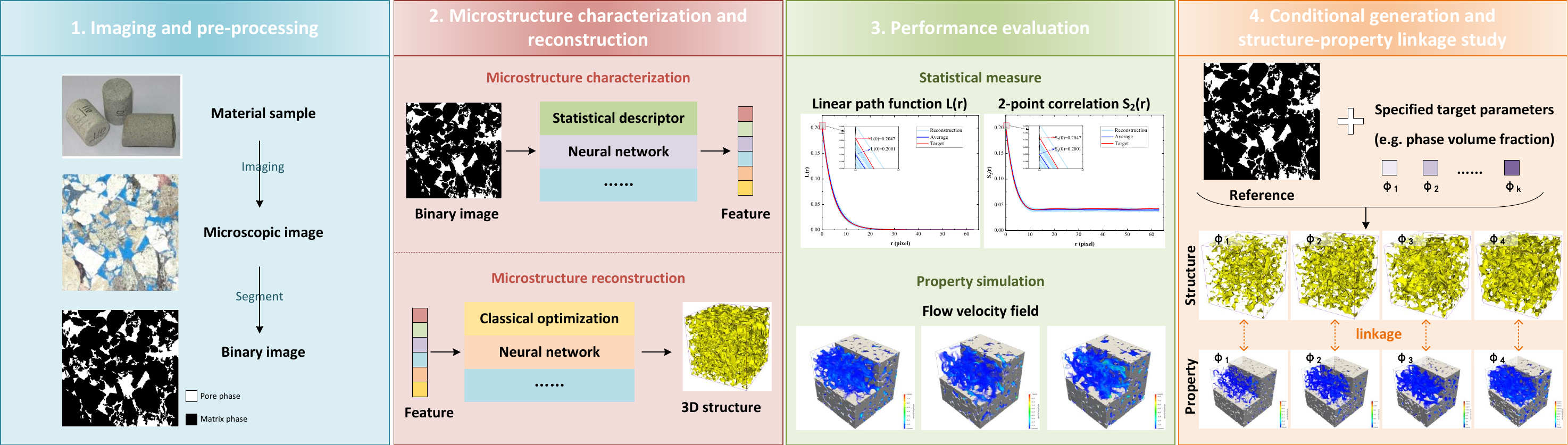}
	\caption{The whole system workflow for structure–property linkage study, including (1) microscopic imaging and pre-processing, (2) microstructure characterization and reconstruction, (3) performance evaluation, and (4) conditional generation and linkage study.}
	\label{fig:0}
\end{figure}

Typically, the entire system workflow (Figure \ref{fig:0}) consists of four main parts: (1) microscopic imaging and pre-processing, (2) microstructure characterization and reconstruction, (3) performance evaluation, and (4) conditional generation and linkage study. The process begins with obtaining microscopic images of the material sample using imaging techniques such as optical microscopy. These acquired images undergo pre-processing steps, including denoising and segmentation \citep{JOBE3-kim2022gradient}, to produce segmented images. Next, microstructure characterization and reconstruction technologies are employed to extract features from the segmented images. Based on these extracted features, 3D microstructures are reconstructed. Subsequently, the performance of the generated structures is evaluated through statistical measures and property analysis. Finally, given reference images and specified target properties (e.g., phase volume fraction), microstructure characterization and reconstruction technologies are used to generate microstructures with the desired properties. The linkage between statistical characteristics and physical properties can be further established through existing and newly generated microstructure samples \citep{10_zang2024psp,21_1kang2024hybrid}. Thus, microstructure characterization and reconstruction methods are critical in these systems and engineering applications, especially for achieving controllable reconstruction. 

Microstructure characterization and reconstruction methods encompass a variety of approaches, including classical optimization methods \citep{12_karsanina2018hierarchical,13_zhang2019efficient,14_xiao2023novel}, texture synthesis methods \citep{15_liu2015random,16_pourfard2017pcto}, random field methods \citep{17_jiang2013efficient,18_gao2021ultra,18_1guo2023spherical}, and machine learning (or deep learning) techniques \citep{19_mosser2017reconstruction,20_feng2020end,21_kench2021generating,EWC25-seibert2024fast}. These methods typically employ statistical descriptors or feature representations to characterize microstructures. Classical optimization methods utilize statistical correlation functions (e.g. 2-point probability functions \citep{22_yeong1998reconstructing} and 2-point cluster functions \citep{23_jiao2014modeling}) as descriptors for microstructural characteristics. Texture synthesis methods \citep{24_xiao2021texture,25_chen20232} model microstructures by representing them as local probability distributions based on texture features, capturing the spatial arrangement of microstructural elements. Machine learning and deep learning methods, such as generative adversarial network-based methods \citep{26_gayon2020pores,27_zhang2023pm,28_argilaga2024synthesising}, leverage neural networks to extract high-level semantic features from microstructures, enabling a more nuanced representation of complex patterns. After extracting statistical or semantic features, microstructure characterization and reconstruction methods proceed to design and implement mapping processes and constraint strategies tailored to generate desired microstructures, such as simulated annealing \citep{29_haghverdi2021modified}, pattern matching \citep{30_yan2023multiscale}, gradient optimization \citep{31_seibert2022descriptor,32_ma2024stochastic}, deep neural networks\citep{32_1fu2021statistical,32_2fu2023hierarchical}, flow-based model \citep{33_guan2021reconstructing}, recurrent neural network \citep{34_zhang20223d}, and diffusion-based model \citep{35_buzzy2024statistically,36_lee2024denoising}.

In recent years, scholars \citep{37_robertson2023local,38_chi2023reconstruction,39_luo2024multi} have made efforts in exploring the characterization of microstructures to achieve controllable generation. With controlling initial phase distributions, \cite{40_karsanina2023stochastic} introduced a method to initialize different phase distributions across various regions within a microstructure. They employed the simulated annealing technique to generate heterogeneous structures characterized by distinct statistical parameter distributions and permeability tensors. \cite{41_xu2022harnessing} constructed a 2D microstructure training dataset using existing image data and stochastically reconstructed images. Through variational autoencoders (VAEs), they learned the latent space distribution and utilized transitions and optimizations within this space to generate stochastic or periodic structures with desired mechanical properties. \cite{42_yang2021exploration} applied a stacked generative adversarial network (StackGAN-v2) to train a neural network model on a piezoceramic image dataset. This allowed them to obtain a latent space representation and explore interpolations within it, resulting in periodic microstructures with varying piezoelectric properties. \cite{43_generale2024inverse} proposed an inverse microstructure design framework that integrates statistical correlation functions with machine learning techniques. Using principal component analysis (PCA) and $\beta$-VAE, they learned the latent space of 2-point spatial correlations. The structure–property linkage was modeled using a multi-output Gaussian regression process. Finally, flow-based generative models and deep variational inference were used to generate weave ceramic matrix composites with targeted thermal conductivities.

Compared to classical optimization methods, machine learning and deep learning approaches offer significant advantages in microstructure characterization due to their use of high-level semantic features rather than traditional statistical descriptors. These advanced features enable more accurate and nuanced representations of complex microstructures. However, the high-level semantic features extracted by neural networks are often difficult to interpret and quantify, posing challenges for achieving explicit controllability. Specifically, when the sample size of microstructure datasets is limited, accurately inferring the latent space of feature distributions becomes particularly challenging. This limitation hinders the direct application of these features for precise and controllable microstructure generation. Therefore, it is critical to explore descriptors that balance accuracy and interpretability while ensuring fast and efficient reconstruction. Achieving statistically controllable generation of microstructures remains a key objective. 

To this end, we propose a novel method that integrates neural networks with sliced-Wasserstein metric (SWNN). This approach leverages local pattern distributions for microstructure characterization, employs local pattern sampling strategies to achieve controllability of global statistical features, and combines neural networks and sliced-Wasserstein metrics for fast and efficient reconstruction. The proposed SWNN method offers several advantages due to its representation of local pattern distributions: (1) Low dataset dependence. The SWNN method remains effective even when the microstructure dataset contains only a few or even a single data sample. It can extract feature distributions and characterize microstructures accurately under such limited conditions. (2) Enhanced interpretability. Local patterns inherently reflect the morphology and detailed information of microstructures. Moreover, the global statistical features of microstructures can be approximately inferred from the combination of all local patterns. Since the statistical features of each local pattern can be directly calculated and quantified, this provides a level of interpretability to the generated microstructures. (3) Controllability. By reproducing similar local patterns with varying distribution compositions (pattern histograms), the SWNN method can generate microstructures with different global features. This capability enables the controllable generation of microstructures with specific global statistical characteristics.

The main contributions of this work are summarized as follows:
\begin{itemize}
    \item The use of local pattern representation and pattern sampling strategies provides precise microstructure characterization and controllable global statistical features. By combining the neural networks with sliced-Wasserstein metric, we achieve accurate, fast, and efficient microstructure reconstruction. 
    \item We have designed 2D-to-3D reconstruction frameworks that are effective even when only a small number of samples or even a single sample is available. These frameworks improve generalization and applicability in scenarios with limited data.
    \item A local–global decoupling reconstruction paradigm has been developed, utilizing spatial location masks to enable the generation of spatially heterogeneous microstructures and complex microstructures.
\end{itemize}

The subsequent sections of this paper are organized as follows: Section \ref{sec2} details our proposed method. Section \ref{sec3} presents the reconstruction experiments and results. Finally, Section \ref{sec4} provides a discussion and Section \ref{sec5} draws a conclusion.

\section{Method}\label{sec2}

\subsection{Problem statement}

Given a microstructure or a set of microstructures $\hat Y$, microstructure characterization and reconstruction methods aim to develop techniques for generating similar microstructures $Y$. The goal is for $Y$ to exhibit characteristics (such as statistical features, morphological structures, and mechanistic properties) that closely resemble those of $\hat Y$. 

For controllable reconstruction, these methods should also be adaptable to specific parameters, enabling the generation of microstructures with targeted characteristics, such as 1-point or 2-point statistics. This presents a challenge in selecting and designing appropriate microstructure descriptors. On one hand, the chosen or designed descriptors should effectively incorporate the specified parameter characteristics. On the other hand, they should be easily adjustable and capable of reproducing these specific parameter characteristics accurately. 

Therefore, as illustrated in Eq. \ref{Eq:1}, controllable microstructure characterization and reconstruction methods involve additional conditional parameters $\eta$. These methods use a control strategy $c_g$ to adjust the features extracted by the descriptor $D$, and employ a mapping process $G$ (as shown in Eq. \ref{Eq:2}) to generate the final reconstructed microstructure $Y$. It is important to note that the final reconstructed microstructure $Y$ should exhibit similar descriptor features to the given microstructure $\hat Y$, while also satisfying the characteristics defined by the given parameter $\eta$. 
\begin{equation}\label{Eq:1}
\mathop {\min }\limits_Y \left\{ {d\left( {D\left( Y \right),{c_g}\left[ {D\left( {\hat Y} \right),\eta } \right]} \right)} \right\}
\end{equation}
where $\hat Y$ represent the given microstructure, $Y$ represents the generated microstructure, $D$ denotes the selected or designed microstructure descriptor, $d$ is the distance metric between two feature distributions, $\eta$ represents the given parameters, and $c_g$ is the control strategy used to adjust the extracted features.

Finally, controllable characterization and reconstruction methods also incorporate an accurate and efficient reconstruction process. This process includes a mapping function $G$ (as defined in Eq. \ref{Eq:2}), which realizes microstructure generation from an initial input $X$ to the final reconstructed microstructure $Y$.
\begin{equation}\label{Eq:2}
G:X \to Y\
\end{equation}

In summary, controllable microstructure characterization and reconstruction methods consist of four key components. (1) Microstructure descriptor $D$, incorporates statistical features. (2) Control strategy $c_g$, generates a feature distribution based on given conditional parameters $\eta$. (3) Distance metric scheme $d$, measures the distance between two feature distributions. (4) Reconstruction process $G$, facilitates rapid and accurate microstructure generation.

Our method specifically adopts local patterns as the microstructure descriptor $D$, modeling the microstructure $\hat Y$ as a local pattern distribution based on Markov Random Field Theory \citep{44_li2009markov}. This descriptor allows us to capture detailed morphological information at the local level. Subsequently, our method extracts statistical information from these local patterns and use it to approximately estimate the global statistical features of the microstructure. By constructing different combinations of local patterns, our control strategy $c_g$ can express the statistical features of microstructures for given specific parameters $\eta$.

To achieve fast and efficient reconstruction process $G$, our method integrates Optimal Transport Theory \citep{45_kolouri2017optimal} with machine learning techniques. To be specific, our method builds neural network models that leverage sliced-Wasserstein metric as the distance metric scheme $d$ to measure the discrepancy between local pattern distributions. This combination ensures both accuracy and efficiency in generating microstructures that closely match the desired characteristics.

\subsection{Microstructure characterization}

\subsubsection{1-point and 2-point statistics}

Based on Statistical Continuum Theory \citep{46_kroner1972statistical} and Strong Contrast Homogenization Theory \citep{47_torquato1997effective,48_torquato1998effective}, a microstructure or a set of microstructures $Y$ can be modeled using an indicator function $I^{(i)}(y)$, as shown in Eq. \ref{Eq:3}. Here, $y$ represents an element at any spatial position within the microstructure $Y$, and $i$ denotes the phase to which this element belongs.
\begin{equation}\label{Eq:3}
{I^{\left( i \right)}}\left( y \right) = \left\{ {\begin{array}{*{20}{c}}
{1,y \in phase{\ }i}\\
{0,y \notin phase{\ }i}
\end{array}} \right.
\end{equation}

Subsequently, using the indicator function $I^{(i)}(y)$, $n$-point probability function $S^{(i)}_n$ \citep{49_torquato2002statistical} can be denoted by Eq. \ref{Eq:4}, where $y_k$ (for $k = 1, 2, \ldots, n$) represents elements at any spatial positions within the microstructure $Y$, and $\langle  \cdot \rangle $ denotes the ensemble average over all possible configurations of the microstructure.
\begin{equation}\label{Eq:4}
S_n^{\left( i \right)}\left( {{y_1},{y_2}, \ldots ,{y_n}} \right) = \langle {I^{\left( i \right)}}\left( {{y_1}} \right){I^{\left( i \right)}}\left( {{y_2}} \right) \cdots {I^{\left( i \right)}}\left( {{y_n}} \right)\rangle 
\end{equation}

$N$-point probability functions can accurately characterize the spatial correlations and higher-order statistical features of microstructures. However, these functions also require more storage and computational resources compared to the microstructure data itself. Consequently, in practice, 1-point statistics (as defined in Eq. \ref{Eq:5}) and 2-point statistics (as defined in Eq. \ref{Eq:6}) are widely used for microstructure characterization due to their computational efficiency. While some researches \citep{50_cheng2022data,51_postnicov2024evaluation} have explored 3-point probability functions, their increased complexity and resource demands limit their practical application. Therefore, 1-point and 2-point statistics remain predominant tools for capturing essential microstructural features while balancing computational feasibility.
\begin{equation}\label{Eq:5}
S_1^{\left( i \right)}\left( y \right) = \langle {I^{\left( i \right)}}\left( y \right)\rangle  = {\phi ^{\left( i \right)}}
\end{equation}
\begin{equation}\label{Eq:6}
S_2^{\left( i \right)}\left( {{y_1},{y_2}} \right) = \langle {I^{\left( i \right)}}\left( {{y_1}} \right){I^{\left( i \right)}}\left( {{y_2}} \right)\rangle 
\end{equation}

1-point statistics can reflect the volume fraction $\phi^{(i)}$ of phase $i$ within the microstructure. For porous material, the volume fraction of the pore phase (porosity) is a critical parameter closely related to most macroscopic physical properties. When the microstructure satisfies the assumptions of stationarity or homogeneity \citep{49_torquato2002statistical}, 2-point statistics depend only on the relative position $r$ between two points, not on their absolute positions. Mathematically, this relationship can be expressed as $S_2^{\left( i \right)}\left( {{y_1},{y_2}} \right) = S_2^{\left( i \right)}\left( r \right),r = \left\| {{y_2} - {y_1}} \right\|$. Similar to 1-point statistics, 2-point statistics also provide information about the volume fraction of phase $i$, denoted as ${\phi ^{\left( i \right)}} = S_2^{\left( i \right)}\left( {r = 0} \right)$.

Moreover, 2-point statistics offer insights into important parameters such as specific surface area and autocorrelation distance \citep{22_yeong1998reconstructing}, which are crucial for understanding macroscopic physical properties. Consequently, numerous studies have focused on predicting these macroscopic physical properties based on 2-point statistics, include effective conductivity \citep{52_li2006quantitative}, mechanical properties \citep{53_belvin2009application}, fluid permeability \citep{54_roding2017functional}, and thermal conductivity \citep{43_generale2024inverse}. Therefore, leveraging the relationship between 1-point statistics, 2-point statistics, and physical properties, it is possible to generate microstructures with specific characteristics by controlling these statistical measures, serving as a powerful tool for integrated computational materials engineering and structure–property linkage studies.

\subsubsection{Local pattern distribution representation}

Notably, the use of 1-point and 2-point statistics can effectively characterize the spatial correlations and structural features of microstructures and offers a degree of controllability. However, these statistical measures have limitations in expressing local morphological details and finer structural features. To address these limitations, local pattern distribution representation is introduced to provide a more detailed characterization of microstructural details. This descriptor enhances the ability to capture and control local morphologies, thereby offering a more comprehensive description of microstructures.

For a microstructure or a set of microstructures $Y$, there exists a probability density distribution $P(Y)$. Consequently, a specific microstructure $Y$ can be viewed as a realization obtained by sampling from this distribution $P(Y)$. However, dealing with such a high-dimensional probability density distribution presents some challenges. Accurate inference and modeling require a substantial number of microstructure samples, which can be resource-intensive. Moreover, inferring the properties of such a high-dimensional distribution is inherently complex and often intractable.

Fortunately, based on Markov Random Field Theory \citep{44_li2009markov}, the microstructure $Y$ can be further modeled as a representation of the local pattern distribution $P(t)$, as shown in Eq. \ref{Eq:7}. Here, $D$ denotes the descriptor used to represent the local pattern distribution, and $t$ represents the local patterns extracted from the microstructure $Y$.
\begin{equation}\label{Eq:7}
D\left( Y \right) = \left\{ {t_1^{\left( Y \right)},t_2^{\left( Y \right)}, \ldots ,t_N^{\left( Y \right)}} \right\} \to P\left( t \right)
\end{equation}

In this context, the microstructure $Y$ can be regarded as a joint expression of local patterns $\left\{ {t_1^{\left( Y \right)},t_2^{\left( Y \right)}, \ldots ,t_N^{\left( Y \right)}} \right\}$. The local pattern $t$ reflects the local morphology and detailed features of the microstructure $Y$, while the local pattern distribution $P(t)$ captures the global features of $Y$ to a certain extent. Specifically, the local pattern $t$ serves as an explicit representation of the local detail information within the microstructure $Y$. Importantly, it also encompasses local 1-point and 2-point statistics. The global 1-point and 2-point statistics (where the correlation distance is less than half the size of the local pattern) of the microstructure $Y$ can be approximately estimated by averaging all local 1-point and 2-point statistics, as shown in Eq. \ref{Eq:8}.
\begin{equation}\label{Eq:8}
S_2^{\left( i \right)\left( Y \right)}\left( r \right) \approx \frac{1}{N}\sum\limits_{j = 1}^N {S_2^{\left( i \right)\left( {{t_j}} \right)}\left( r \right)} ,0 \le r \le \frac{l_t}{2}
\end{equation}
where $S_2^{(Y)}$ and $S_2^{(t)}$ represent the 2-point statistics of the microstructure $Y$ and the local pattern $t$, $N$ represents the total number of local patterns within the microstructure, and $l_t$ denotes the size of the local pattern.

More importantly, this representation of local pattern distributions enables the modeling of microstructures even when only a few microstructure samples or even a single sample are available. By reproducing the same or similar representations of local pattern distributions, it becomes possible to generate microstructures that exhibit identical statistical characteristics and analogous morphological mechanisms.

\subsubsection{Control strategy}

To achieve controllable microstructure characterization, two primary objectives should be satisfied. (1) Statistical consistency and controllability. The generated microstructure should satisfy the given conditional parameters. Specifically, Eq. \ref{Eq:9} ensures that for any given conditional parameter, the 2-point statistics of the reconstructed microstructure Y match those specified $\eta$. (2) Morphological similarity. The generated microstructure should exhibit morphological characteristics similar to those of the given reference microstructure. Eq. \ref{Eq:10} ensures that for any local pattern $t$ in the reconstructed microstructure $Y$, there are corresponding similar local patterns in the reference microstructure $\hat Y$.
\begin{equation}\label{Eq:9}
\forall \eta  \in \left[ {0,1} \right],{\ }\exists Y,{\ }s.t.{\ }S_2^{\left( i \right)\left( Y \right)}\left( r \right) = \eta
\end{equation}
\begin{equation}\label{Eq:10}
\forall {t_p} \in Y,{\ }\exists {t_q} \in \hat Y,{\ }s.t.{\ }{t_p} = {t_q}
\end{equation}

Our method leverages local patterns to characterize the microstructure, offering several advantages in meeting both objectives. One advantage of using local pattern distribution representation is its intuitive and interpretable nature. The extracted local patterns directly reflect the morphological characteristics of the microstructure. Thus, accurately reproducing these local patterns from the reference microstructure ensures the consistency of morphological features. Another advantage is the enhanced controllability offered by local pattern distribution representation. By varying the combination of local patterns, different local pattern distributions can be achieved, each representing distinct statistical features. This flexibility allows for the microstructure reconstruction with controllable global statistics.

Therefore, given a set of parameters $\eta$ and a local pattern distribution $D(Y)$ or $P(t)$ characterizing the microstructure $Y$, the control strategy $c_g$ aims to generate a new local pattern distribution $P_{\eta}(t)$ that satisfies the specified parameters $\eta$. This is achieved by purposefully sampling from the original local pattern distribution $D(Y)$ or $P(t)$, as shown in Eq. \ref{Eq:11}.
\begin{equation}\label{Eq:11}
{c_g}\left[ {D\left( Y \right),\eta } \right] = \left\{ {t_{s1}^{\left( Y \right)},t_{s2}^{\left( Y \right)}, \ldots ,t_{sN}^{\left( Y \right)}} \right\} \to {P_\eta }\left( t \right)
\end{equation}
where $\left\{ {t_{s1}^{\left( Y \right)},t_{s2}^{\left( Y \right)}, \ldots ,t_{sN}^{\left( Y \right)}} \right\}$ is the newly generated combination of local patterns that satisfies $\tfrac{1}
{N}\sum\nolimits_{k = 1}^N {S_2^{\left( i \right)\left( {t_{sk}} \right)} \left( r \right)} = \eta $, $t_{sk}$ represents the local patterns sampled from the original local pattern distribution $D(Y)$ or $P(t)$, and $P_{\eta}(t)$ is the expected distribution that meets the specified parameters $\eta$. 

In previous work, several methods \citep{55_ding2018improved,56_xia2021three} have been developed to achieve the controllability of global statistical parameters by selecting appropriate local patterns during pattern matching-based reconstruction process. Similarly, the goal of the designed control strategy $c_g$ is to sample suitable local patterns and subsequently realize specific combinations of these patterns to achieve precise control over global statistical parameters.

Given a set of parameters $\eta$ and a local pattern distribution $D\left( Y \right) = \left\{ {t_1,t_2, \ldots ,t_N} \right\}$, characterizing the microstructure $Y$, the control strategy $c_g$ is implemented as follows:

\textbf{Step 1.} Calculate 1-point statistics $s_k$ for all local patterns ${t_k}$. Sort the local patterns based on their 1-point statistics calculation results, satisfying $s_{c\left[ k \right]}  \leqslant s_{c\left[ {k + 1} \right]}$.

\textbf{Step 2.} Calculate the average of the 1-point statistics for all local patterns, denoted as $\bar s = \tfrac{1}{N}\sum\nolimits_{k = 1}^N {s_k}$. Compare the average $\bar s$ with the given parameter $\eta$:

\begin{itemize}
    \item If $\bar s \le \eta$ and $s_{c\left[ N \right]} \ge \eta$: There exists a subset of local patterns $\left\{ {t_{c\left[ m \right]} ,t_{c\left[ {m + 1} \right]} , \ldots ,t_{c\left[ N \right]} } \right\}$, ${m} \in \left\{ {1,2, \ldots ,N} \right\}$, such that $\tfrac{1}{{N - m + 1}}\sum\nolimits_{k = m}^N {s_{c\left[ k \right]} }  \leqslant \eta $ and ${\textstyle{1 \over {N - {m}}}}\sum\nolimits_{k = {m} + 1}^N {s_{c\left[ k \right]}}  > \eta$, where $\left\{ {t_{c\left[ m \right]} ,t_{c\left[ {m + 1} \right]} , \ldots ,t_{c\left[ N \right]} } \right\}$ is a combination of local patterns satisfying the preliminary requirements. 
    \item If $\bar s > \eta$ and $s_{c\left[ 1 \right]} \le \eta$: There exists a subset of local patterns $\left\{ {t_{c\left[ 1 \right]} ,t_{c\left[ 2 \right]} , \ldots ,t_{c\left[ m \right]} } \right\}$, ${m} \in \left\{ {1,2, \ldots ,N} \right\}$, such that ${\textstyle{1 \over {m}}}\sum\nolimits_{k = 1}^{m} {s_{c\left[ k \right]}}  \le \eta$ and ${\textstyle{1 \over {{m}+1}}}\sum\nolimits_{k = 1}^{{m}+1} {s_{c\left[ k \right]}}  > \eta$, where $\left\{ {t_{c\left[ 1 \right]} ,t_{c\left[ 2 \right]} , \ldots ,t_{c\left[ m \right]} } \right\}$ is a combination of local patterns satisfying the preliminary requirements.
    \item Otherwise: No desired local pattern combination exists for the current local pattern size.
\end{itemize}

\textbf{Step 3.} Expand the newly obtained local pattern combination ($\left\{ {t_{c\left[ m \right]} ,t_{c\left[ {m + 1} \right]} , \ldots ,t_{c\left[ N \right]} } \right\}$ or $\left\{ {t_{c\left[ 1 \right]} ,t_{c\left[ 2 \right]} , \ldots ,t_{c\left[ m \right]} } \right\}$) to ensure the number of elements matches the number of local patterns in the given distribution $D\left( Y \right)$.

\textbf{Step 4.} Output the expanded local pattern combination as the desired local pattern distribution $\left\{ {t_{s1},t_{s2}, \ldots ,t_{sN}} \right\}$.

In Step 3, the expanding operation aims to make the number of elements in the input distribution $\left\{ {{t_k}} \right\}_{k = 1}^{{m}}$ coincide with the number of elements in the given distribution $\left\{ {{t_k}} \right\}_{k = 1}^N$ while preserving the original distribution as much as possible. This is achieved through integer multiples replication and random sampling for supplement, denoted as $\left\{ {{t_k}} \right\}_{k = 1}^N = \bigcup\nolimits_{{j_c}} {\left\{ {{t_k}} \right\}_{k = 1}^{{m}}}  \cup \left\{ {{t_k}} \right\}_{k = 1}^{j_r}$, where $j_c  = \left\lfloor {{N \mathord{\left/
 {\vphantom {N m}} \right.
 \kern-\nulldelimiterspace} m}} \right\rfloor $ is the integer multiple used for replication and $j_r  = N - j_c  \times m$ represents the additional patterns randomly sampled from the original distribution. The original distribution is replicated by integer multiples to closely match the size of the given distribution. This ensures that the core characteristics of the original distribution are preserved. If there are still insufficient elements after replication, the remaining elements are supplemented by randomly sampling from the original distribution. This process ensures that the expanded distribution remains consistent with the original pattern while achieving the desired size.

Through this control strategy $c_g$, we can obtain a representation of specific microstructures $\left\{ {t_{s1},t_{s2}, \ldots ,t_{sN}} \right\}$ that exhibit similar local details but different global statistical characteristics. Importantly, this strategy allows for generating diverse microstructure representations using only a single microstructure sample, and enhances the flexibility and applicability of microstructure modeling, enabling exploration and design based on limited data.

\subsection{Microstructure reconstruction}

\subsubsection{Neural network-based reconstruction}

Through the local pattern distribution representation and the control strategy, our method achieves an accurate and controllable representation of the microstructure. In general, this representation can be treated as an optimization objective for some descriptor-based gradient optimization methods. The process typically involves randomly initializing the generated microstructure and iteratively updating it using gradient optimization methods to obtain the final generated microstructure. While combining local pattern distribution representation with gradient optimization reconstruction offers high efficiency for single reconstructions, there are limitations when acquiring a large number of microstructure samples. Additionally, for large-size microstructures (e.g., 512-size or 1024-size), direct optimization methods consume significant video memory, especially on GPU-based systems. Running these methods on CPUs results in reduced reconstruction efficiency due to limited computational resources.

To address the challenges associated with multiple reconstructions and large-size microstructure generation, our method incorporates neural networks to enhance efficiency. Instead of relying on iterative optimization processes for generating microstructures, our method employs neural networks to establish this mapping. To be specific, we aim to construct a mapping from the source domain $X \sim {\rm N}\left( {\mu ,\sigma} \right)$ (Gaussian distribution) to the target domain $Y \sim P\left( t \right)$ using neural networks, as shown in Eq. \ref{Eq:2}. Here, $P(t)$ represents the local pattern distribution, which is approximately equivalent to the assumed microstructure distribution $P(Y)$ under the Markov random field assumption.

In addition, the trained neural network model is expected to exhibit generalization capabilities, allowing it to generate spatially heterogeneous microstructures by controlling the input distribution. Therefore, the designed neural network model is expected to satisfy the following three conditions. (1) The input size of the neural network should closely match the sizes of the given microstructure. This allows for the introduction of spatial location masks, enabling controllability over spatial locations and facilitating the generation of spatially heterogeneous microstructures. (2) The neural network architecture should be simple yet effective. Given that the input size is similar to that of the given microstructure, the training process can incur memory overhead. A simpler structure helps mitigate the storage and computational demands associated with processing intermediate results, thereby reducing video memory consumption. (3) The loss function should accurately measure the distance between the local pattern distributions. This ensures the stability of both the training and reconstruction processes and guarantees the accuracy of the generated microstructure.

\subsubsection{Neural network structure}

Considering these factors, our method adopts a network structure similar to that described in literature \citep{57_ma2023fast} to construct the mapping. The chosen architecture is simple yet effective, and its design can be flexibly adjusted based on the autocorrelation distance of the microstructure. The entire network consists of $l_m$ BlockA modules followed by a single BlockB module. The value of $l_m$ is determined based on the autocorrelation distance $l_c$ and the receptive fields of BlockA and BlockB ($R_A$ and $R_B$), as detailed in Eq. \ref{Eq:12}.
\begin{equation}\label{Eq:12}
l_m = \left\lceil {{{\left( {{l_c} - {R_B} - 1} \right)} \mathord{\left/
 {\vphantom {{\left( {{l_c} - {R_B} - 1} \right)} {\left( {{R_A} - 1} \right)}}} \right.
 \kern-\nulldelimiterspace} {\left( {{R_A} - 1} \right)}}} \right\rceil
\end{equation}

Specifically, both BlockA and BlockB are composed of three key layers: a convolutional layer, an activation layer, and a batch normalization (BN) layer. The input and output calculations for each of these layers are detailed in Eq. \ref{Eq:13}.
\begin{equation}\label{Eq:13}
\left\{ {\begin{array}{*{20}l}
   {x^{\left( l \right)}  = \theta _w^{\left( l \right)} x^{\left( {l - 1} \right)}  + \theta _b^{\left( l \right)} } \hfill  \\
   {h\left( {x^{\left( l \right)} } \right) = \max \left( {0,x^{\left( l \right)} } \right) + \min \left( {0,ax^{\left( l \right)} } \right)} \hfill  \\
   {x^{\left( l \right)}  = {{\gamma \left( {x^{\left( l \right)}  - \mu _x } \right)} \mathord{\left/
 {\vphantom {{\gamma \left( {x^{\left( l \right)}  - \mu _x } \right)} {\sigma _x }}} \right.
 \kern-\nulldelimiterspace} {\sigma _x }} + \beta  \equiv BN_{\gamma ,\beta } \left( {x^{\left( l \right)} } \right)} \hfill  
 \end{array} } \right.
\end{equation}
where ${\theta}_{w}$ and ${\theta}_{b}$ are are convolutional layer parameters, $h\left( x \right)$ is the LeakyRelu activation function, ${\mu}_x$ and ${\sigma}_x$ are the mean and standard deviation of the input, $\gamma$ and $\beta$ are the scale and shift parameters of the BN layer.

For the convolutional layers within BlockA, a 3-size convolution kernel is used, while for the convolutional layers within BlockB, a 1-size convolution kernel is employed. Consequently, the input and output transformations through BlockA and BlockB can be further denoted as shown in Eq. \ref{Eq:14}.
\begin{equation}\label{Eq:14}
\left\{ {\begin{array}{*{20}{l}}
{\text{BlockA}:{x^{\left( l \right)}} = {g_1}\left( {{x^{\left( {l - 1} \right)}};\theta } \right) = BN\left( {h\left( {\theta _{{w_1}}^{\left( l \right)}{x^{\left( {l - 1} \right)}} + \theta _{{b_1}}^{\left( l \right)}} \right)} \right)}\\
{\text{BlockB}:{x^{\left( l \right)}} = {g_2}\left( {{x^{\left( {l - 1} \right)}};\theta } \right) = BN\left( {h\left( {\theta _{{w_2}}^{\left( l \right)}{x^{\left( {l - 1} \right)}} + \theta _{{b_2}}^{\left( l \right)}} \right)} \right)}
\end{array}} \right.
\end{equation}

Finally, using the neural network model composed of $l_m$ BlockA modules followed by 1 BlockB module, any input structure $X$ sampled from a Gaussian distribution can be transformed into a generated microstructure $Y$. This transformation can be mathematically expressed as shown in Eq. \ref{Eq:15}, where $X$ represents the input structure sampled from a Gaussian distribution ${\rm N}\left( {\mu ,\sigma } \right)$, $Y$ denotes the generated microstructure, $G$ is the mapping function implemented by the neural network, and $\theta$ represents the parameters of the neural network.
\begin{equation}\label{Eq:15}
Y = G\left( {X;\theta } \right) = \left( {{g_1}^{l_m} \circ {g_2}} \right)\left( {X;\theta } \right)
\end{equation}

\subsubsection{Distance measurement scheme}

An accurate and effective distance measurement scheme is critical for comparing high-dimensional probability distributions. Inspired by Optimal Transport Theory \citep{45_kolouri2017optimal}, our method introduces sliced-Wasserstein metric to measure the distance between local pattern distributions of microstructures.

Sliced-Wasserstein metric is a distribution measurement technique designed for high-dimensional probability distributions. It calculates the distance between such distributions by projecting them onto one-dimensional spaces and measuring the distances in these lower-dimensional projections across multiple directions. This approach simplifies the comparison of complex, high-dimensional distributions. Mathematically, sliced-Wasserstein metric between two high-dimensional distributions $U_d$ and $V_d$ is denoted as shown in Eq. \ref{Eq:16}.
\begin{equation}\label{Eq:16}
S{W_p}\left( {{U_d},{V_d}} \right) = {\left( {\int_{{^{d - 1}}} {W_p^p\left( {\left( {\Re {f_\mu }} \right)\left( { \cdot ,\psi } \right),\left( {\Re {f_\nu }} \right)\left( { \cdot ,\psi } \right)} \right)d\psi } } \right)^{\frac{1}{p}}}
\end{equation}
where $ {\Re {f}}$ represents the Radon transform \citep{46_kroner1972statistical} used for one-dimensional projections, denoted as $\left( {\Re f} \right)\left( {z,\psi } \right): = \int_{{\mathbb{R}^d}} {f(\nu )\delta (z - \langle \nu ,\psi \rangle )d\nu }$, and $\psi$ is the projection direction. While $W_p$ represents Wasserstein distance, a metric for one-dimensional distributions, denoted as $W_p^p \left( {U_1 ,V_1 } \right): = \inf _{\gamma  \in \Gamma \left( {U_1 ,V_1 } \right)} \int_{\mathbb{R}^d  \times \mathbb{R}^d } {\left\| {\mu  - \nu } \right\|_p^p d\gamma (\mu ,\nu )} $. Here, $U_1$ and $V_1$ are two one-dimensional distributions, and $\gamma  \in \Gamma \left( {U_1 ,V_1 } \right)$ is the transportation scheme for all plans.

As a distance metric scheme in high-dimensional space, sliced-Wasserstein metric offers two primary advantages. (1) High-dimensional distributions, which are challenging to measure directly, are projected into one-dimensional spaces. This projection allows the distance between high-dimensional distributions to be effectively measured using one-dimensional metrics. By reducing the complexity from high dimensions to one dimension, sliced-Wasserstein metric simplifies the comparison of complex distributions. (2) The distance metric between the projected one-dimensional distributions utilizes Wasserstein metric. Importantly, this measurement has a closed-form solution in practice, which can be efficiently computed by sorting the one-dimensional distributions (as shown in Eq. \ref{Eq:17}). This feature significantly enhances computational efficiency and feasibility.
\begin{equation}\label{Eq:17}
{W_p}\left( {{{\tilde U}_1},{{\tilde V}_1}} \right) = {\left( {\frac{1}{N}\sum\limits_{n = 1}^N {\left\| {{\mu _{{c_\mu }\left[ n \right]}} - {\nu _{{c_\nu }\left[ n \right]}}} \right\|_p^p} } \right)^{\frac{1}{p}}}
\end{equation}
where ${\tilde U_1} = \left\{ {{\mu _k}} \right\}_{k = 1}^N$ and ${\tilde V_1} = \left\{ {{\nu _k}} \right\}_{k = 1}^N$ represent two one-dimensional distributions containing $N$ actual observation samples each. $\left\{ {{\mu _{{c }\left[ k \right]}}} \right\}_{k = 1}^N$ and $\left\{ {{\nu _{{c }\left[ k \right]}}} \right\}_{k = 1}^N$ denote the sorted versions of these one-dimensional distributions.

Therefore, given the actual observed high-dimensional distributions ${\tilde U_d} = \left\{ {{\mu _k}} \right\}_{k = 1}^N$ and ${\tilde V_d} = \left\{ {{\nu _k}} \right\}_{k = 1}^N$, each containing $N$ observation samples, sliced-Wasserstein metric can be approximately estimated using $L$ projection directions. This approximation is detailed in Eq. \ref{Eq:18}.
\begin{equation}\label{Eq:18}
\widetilde {S{W_p}}\left( {{{\tilde U}_d},{{\tilde V}_d}} \right) = {\left( {\frac{1}{{NL}}\sum\limits_{l = 1}^L {\sum\limits_{n = 1}^N {\left\| {\langle {\mu _{c_\mu ^l\left[ n \right]}},{\psi _l}\rangle  - \langle {\nu _{c_\nu ^l\left[ n \right]}},{\psi _l}\rangle } \right\|_p^p} } } \right)^{\frac{1}{p}}}
\end{equation}

\subsection{Framework and implement scheme}

\begin{figure}%[htbp]
	\centering
        \includegraphics[scale=.6]{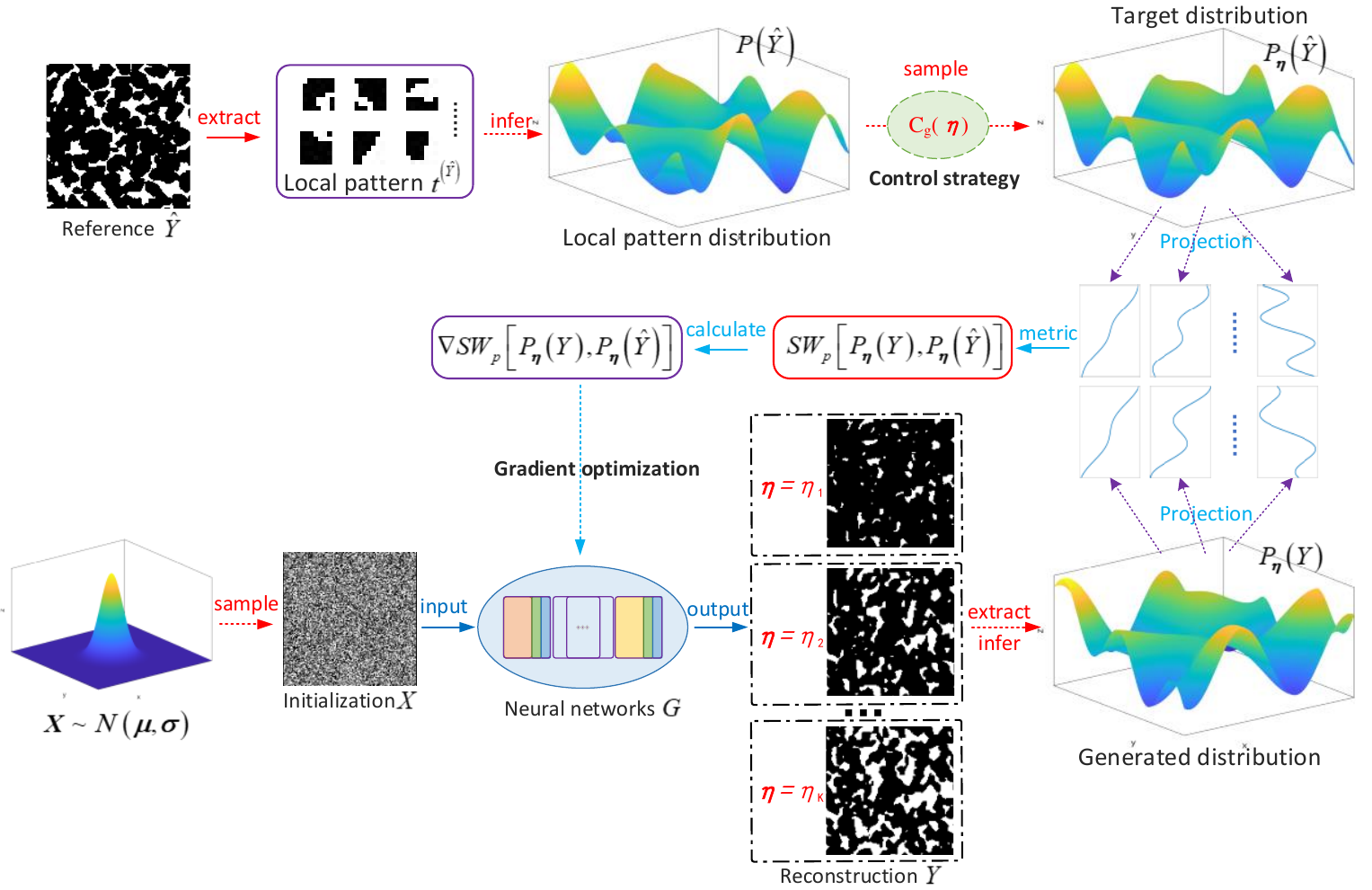}
	\caption{The framework of statistically controllable microstructure reconstruction.}
	\label{fig:2-4-1}
\end{figure}

Based on the aforementioned principles and methods, the controllable characterization and reconstruction framework is illustrated in Figure \ref{fig:2-4-1}. Within this framework, the local pattern distribution is used as the descriptor $D$ to characterize the microstructure, the control strategy $c_g$ is designed to generate a target distribution that aligns with specified conditional parameters $\eta$, sliced-Wasserstein metric serves as the metric scheme $d$ to calculate the distance between the local pattern distribution of the generated microstructure and the target distribution, and combination of neural networks with gradient optimization techniques find a mapping $G$ that transforms the input into the reconstructed microstructure efficiently.

The optimization objective of this framework is presented in Eq. \ref{Eq:19} and \ref{Eq:20}, where $P_{\text{data}}$ is a Gaussian distribution ${\rm N}\left( {\mu ,\sigma } \right)$. The goal is to find a neural network mapping parameterized by $\theta$ such that the generated microstructure $Y$ exhibits local morphology similar to the reference microstructure $\hat Y$ while satisfying specified conditional parameters $\eta$. For 2D reconstructions (Eq. \ref{Eq:19}), our method aims to minimize the difference between the local pattern distribution of the generated microstructure and the target distribution. While for 3D reconstructions (Eq. \ref{Eq:20}), our method seeks to minimize the differences between the local pattern distribution of random sections and the target distribution across each of the three principal directions. In this context, the target microstructure is assumed to satisfy the stationarity or homogeneity assumption. Finally, through the gradient optimization method \citep{58_DBLP:journals/corr/KingmaB14}, as shown in Eq. \ref{Eq:21}, the network parameters are iteratively updated over multiple epochs to obtain a neural network model that meets the optimization objective. The pseudo-code of the training and reconstruction process is shown in Alg \ref{alg:Opt}.
\begin{equation}\label{Eq:19}
\mathop {\min }\limits_\theta  {{\cal L}_{2D}}\left( \theta  \right) = {\mathbb{E}_{X \sim {P_{{\rm{data}}}}}}\left[ {S{W_p}\left( {D\left( {G\left( {X;\theta } \right)} \right),{c_g}\left[ {D\left( {\hat Y} \right),\eta } \right]} \right)} \right]
\end{equation}
\begin{equation}\label{Eq:20}
\mathop {\min }\limits_\theta  {\mathcal{L}_{3D}}\left( \theta  \right) = {\mathbb{E}_{X \sim {P_{{\text{data}}}}}}\left[ {\sum\limits_{k = \text{x}, \text{y}, \text{z}} {S{W_p}\left( {D\left( {G\left( {{X^{\left( k \right)}};\theta } \right)} \right),{c_g}\left[ {D\left( {\hat Y} \right),\eta } \right]} \right)} } \right]
\end{equation}
\begin{equation}\label{Eq:21}
{\theta _{j + 1}} = {\theta _j} - \alpha \nabla \mathcal{L}\left( \theta  \right)
\end{equation}
where ${\theta}_j$ represents the network parameters at iteration $j$, $\alpha$ is the learning rate, and $\nabla \mathcal{L}\left( \theta  \right)$ denotes the gradient of the loss function with respect to the parameters $\theta$.

For large-size microstructure reconstructions, our method employs a chunking strategy since no boundary filling condition is applied. For microstructure generation with a specified reconstruction size, the input is sampled from a Gaussian distribution, and the sampled input is divided into smaller sub-blocks according to an appropriate block size. Subsequently, each sub-block is reconstructed separately using the neural network. Since no boundary filling conditions are required, the neural network processes each sub-block independently, ensuring seamless integration of reconstructed sub-blocks without continuity issues at the spliced boundaries.

The implementation scheme for the entire process is as follows:

\textbf{Step 1.} Extract local pattern distribution. Given a reference microstructure or a set of reference microstructures $\hat Y$, the local pattern distribution is extracted using a predefined template. This extracted distribution is denoted as $D(\hat Y)$.

\textbf{Step 2.} Generate target distribution. Using the control strategy $c_g$, a new local pattern distribution that satisfies the given conditional parameters $\eta$ is sampled from $D(\hat Y)$. This target distribution serves as the optimization objective and is denoted as ${{c_g}\left[ {D\left( {\hat Y} \right),\eta } \right]}$.

\textbf{Step 3.} Mapping process and loss calculation. For each iteration of 2D reconstructions, a 2D initialization input $X$ is sampled from a Gaussian distribution and fed into the neural network to obtain the reconstructed microstructure $Y$; The current local pattern distribution of the reconstructed microstructure $Y$ is then extracted using the same template, denoted as $D(Y)$. The loss function between the current distribution $D(Y)$ and the target distribution ${{c_g}\left[ {D\left( {\hat Y} \right),\eta } \right]}$ is calculated using sliced-Wasserstein metric. For each iteration of 3D reconstructions, three orthogonal 2D initialization inputs $X^{(k)}$ (for $k = \text{x}, \text{y}, \text{z}$) are sampled from a Gaussian distribution and fed into the neural network to obtain three reconstructed sections $Y^{(k)}$ (for $k = \text{x}, \text{y}, \text{z}$). The current local pattern distribution of three sections is then extracted using the same template, denoted as $D(Y^{(k)})$. The average loss function between these distributions and the target distribution ${{c_g}\left[ {D\left( {\hat Y} \right),\eta } \right]}$ is calculated using sliced-Wasserstein metric.

\textbf{Step 4.} Optimization and parameter update. Gradient optimization method is employed to update the parameters of the neural network $\theta$. This iterative process aims to minimize the loss function, thereby obtaining a mapping $G$ that meets the specified requirements.

\textbf{Step 5.} Reconstruction process. an input with the specified size is sampled from a Gaussian distribution. This input is sent to the trained neural network in blocks according to the set block size. The reconstruction results are then integrate and output.

\begin{algorithm}[h]  
	\caption{Training and reconstruction process for statistically controllable reconstruction}  
	\label{alg:Opt}  
	\begin{algorithmic}[1]  
		\Require  
		Reference microstructure $\hat Y$; Conditional parameter $\eta$; Number of optimization iterations $N_{iter}$; Number of microstructure reconstructions $N_c$ of the specified size;
		\Ensure 
		Neural network model parameters $\theta$; Microstructure reconstruction results $\left\{ {Y_i } \right\}_{i = 1}^{N_c }$ ;
		\\Use the predefined template to extract the local pattern distribution $D(\hat Y)$ from the reference microstructure $\hat Y$.
            \State Use the control strategy $c_g$ to generate the target distribution ${{c_g}\left[ {D\left( {\hat Y} \right),\eta } \right]}$ satisfying the conditional parameter $\eta$.
            \State Construct the neural network model and initialize the network parameters $\theta$.
            \Repeat
            \State Let $j \leftarrow 0$.
            \State Initialize the input $X$ (or $X^{\left( k \right)}$, $k=\text{x, y, z}$) by sampling from a Gaussian distribution.
            \State Feed $X$ (or $X^{\left( k \right)}$) into the neural network model to obtain the generated microstructure $Y$ (or $Y^{\left( k \right)}$) based on Eq. \ref{Eq:15}.
            \State Use the same template to extract the local pattern distribution $D(Y)$ (or $D(Y^{\left( k \right)})$) from the generated microstructure $Y$ (or $Y^{\left( k \right)}$).
            \State Calculate the sliced-Wasserstein metric loss $\mathcal{L}\left( \theta  \right)$ between the local pattern distribution $D(Y)$ (or $D(Y^{\left( k \right)})$) and the target distribution ${{c_g}\left[ {D\left( {\hat Y} \right),\eta } \right]}$ based on Eq. \ref{Eq:18} (or Eq. \ref{Eq:19}).
            \State Calculate the backpropagation gradient $\nabla \mathcal{L}\left( \theta  \right)$ and update the network parameters $\theta$ based on Eq. \ref{Eq:21}.
            \State Let $j \leftarrow j+1$. 
            \Until{$j=N_{iter}$}         
            \State Output the neural network model parameters $\theta$.
            \Repeat
            \State Let $i \leftarrow 0$.
            \State Initialize the input $X_i$ of a specified size by sampling from a Gaussian distribution.
            \State Use the chunking strategy to divide the input $X_i$ into the input sub-block sequence.
            \State Feed the input sub-block separately into the trained neural network model to obtain the reconstructed sub-block sequence.
            \State Concatenate the reconstructed sub-blocks to obtain the generated microstructure $Y_i$.
            \State Let $i \leftarrow i+1$. 
            \Until{$i=N_c$}
            \State Output the generated microstructures $\left\{ {Y_i } \right\}_{i = 1}^{N_c }$.
	\end{algorithmic}  
\end{algorithm}

Sample body text. Sample body text. Sample body text. Sample body text. Sample body text. Sample body text. Sample body text. Sample body text.

\section{Experiments and results}\label{sec3}

To verify the effectiveness of our method, reconstruction experiments were conducted on four materials: silica material \citep{59_bostanabad2016characterization}, porous rock \citep{60_coker1996morphology}, battery electrode material \citep{61_feng2018reconstruction}, and superalloy \citep{37_robertson2023local}. The experiments were designed to evaluate various aspects of our method's capabilities. Initially, stochastic reconstructions without conditional parameters were performed to assess the method's ability to characterize and reconstruct microstructures accurately. Next, conditional parameters and control strategies were introduced, and controllable reconstructions were carried out. This verified the method's capability to generate microstructures that satisfy specified conditional parameters, demonstrating its flexibility and adaptability. Subsequently, spatial location masks were incorporated to perform heterogeneous reconstructions. This evaluated the method's ability to generate spatially heterogeneous microstructures, showcasing its versatility in handling complex structural variations. Finally, large-size inputs were initialized and reconstructed in blocks to validate the method's capability to generate extensive microstructures. Experiments included reconstructions of 512- and 1024-size microstructures, highlighting the scalability and efficiency of our method.

\subsection{Stochastic reconstruction without conditional parameters}

To validate the capability of our method for microstructure characterization and reconstruction, we performed 2D-to-3D stochastic reconstructions without conditional parameters on reference images of four materials—silica material, porous rock, battery electrode material, and superalloy. The parameter settings are detailed as follows: number of BlockA modules is set to 8, number of iterations is set to 2000, learning rate is set to 0.05, predefined template sizes for local pattern extraction are set to 5, 7, 6, and 5 for silica material, porous rock, battery electrode material, and superalloy, respectively.

\begin{figure}%[htbp]
	\centering
        \includegraphics[scale=.15]{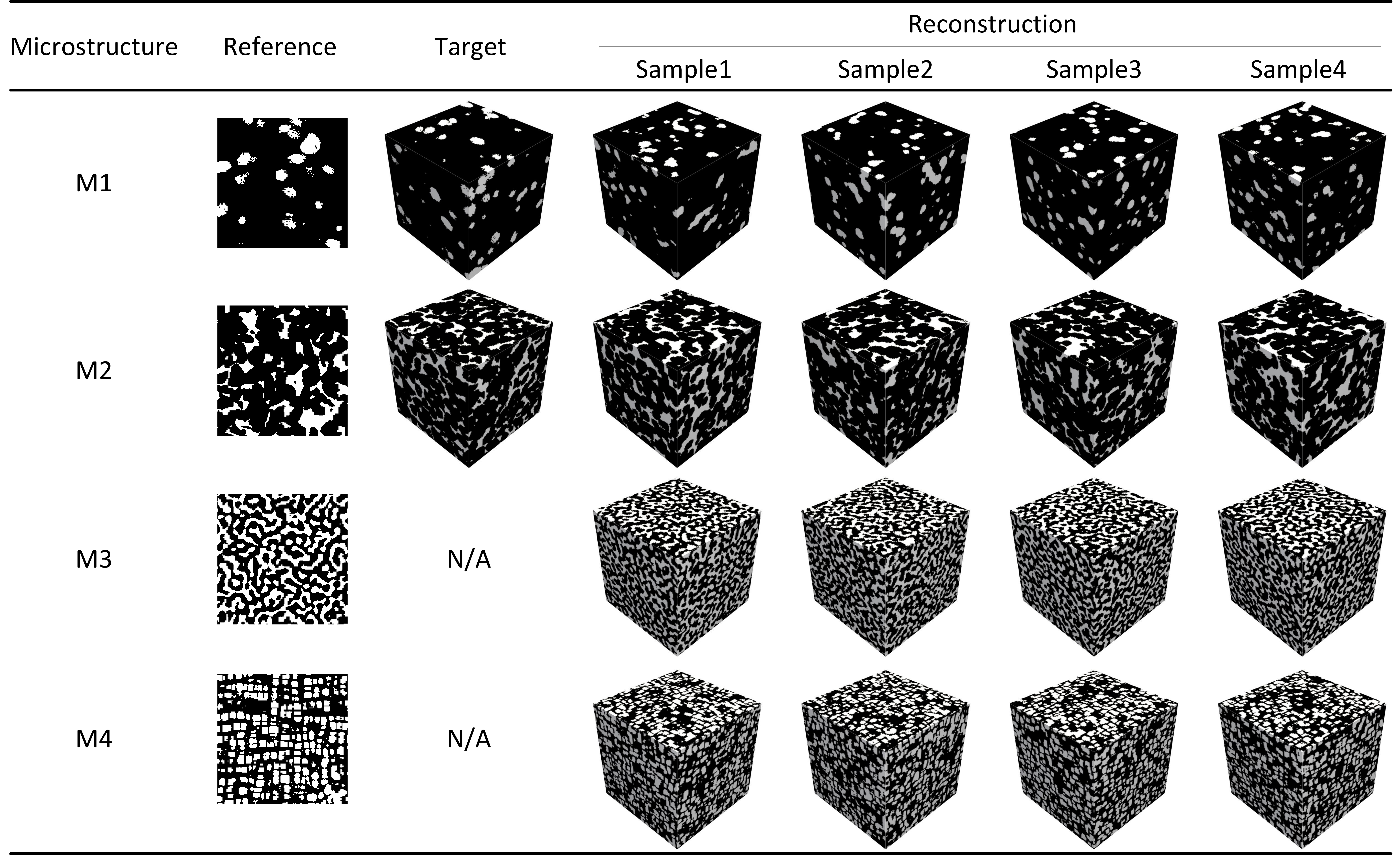}
	\caption{Visual comparisons among reference images, target microstructures and partial reconstruction results. Microstructures M1–M4 are silica material, porous rock, battery electrode material, and superalloy, respectively.}
	\label{fig:3-1-1}
\end{figure}

To ensure the stability and robustness of our method, 4 models were trained separately on reference images of each material, and each trained model was used to perform 5 reconstructions, with a total of 20 sets of reconstruction results per material. Visual comparisons among the reference images, target microstructures, and partial reconstruction results in 2D-to-3D stochastic reconstruction are illustrated in Figure \ref{fig:3-1-1}. The microstructures M1–M4 correspond to silica material, porous rock, battery electrode material, and superalloy, respectively, each with a size of 128.

In the visual comparison presented in Figure \ref{fig:3-1-1}, our reconstruction results effectively reproduce both the overall spatial structure and local morphology of these four materials. Notably, the sections of the reconstruction samples in all three directions exhibit excellent agreement with the reference images and closely resemble the corresponding sections of the target microstructures.

Statistical correlation functions, such as 2-point probability functions and linear path functions \citep{22_yeong1998reconstructing}, can effectively reflect the statistical features and spatial correlations of microstructures. To further measure the accuracy of our reconstruction results, we conducted statistical analyses using these functions. A comparison of linear path functions is provided in Figure \ref{fig:A1} of the Appendix. Additionally, we enlarged the local area at the starting point to clearly highlight the differences between the reconstruction results and the target structures (or reference images).

\begin{figure}%[htbp]
	\centering
        \includegraphics[scale=.20]{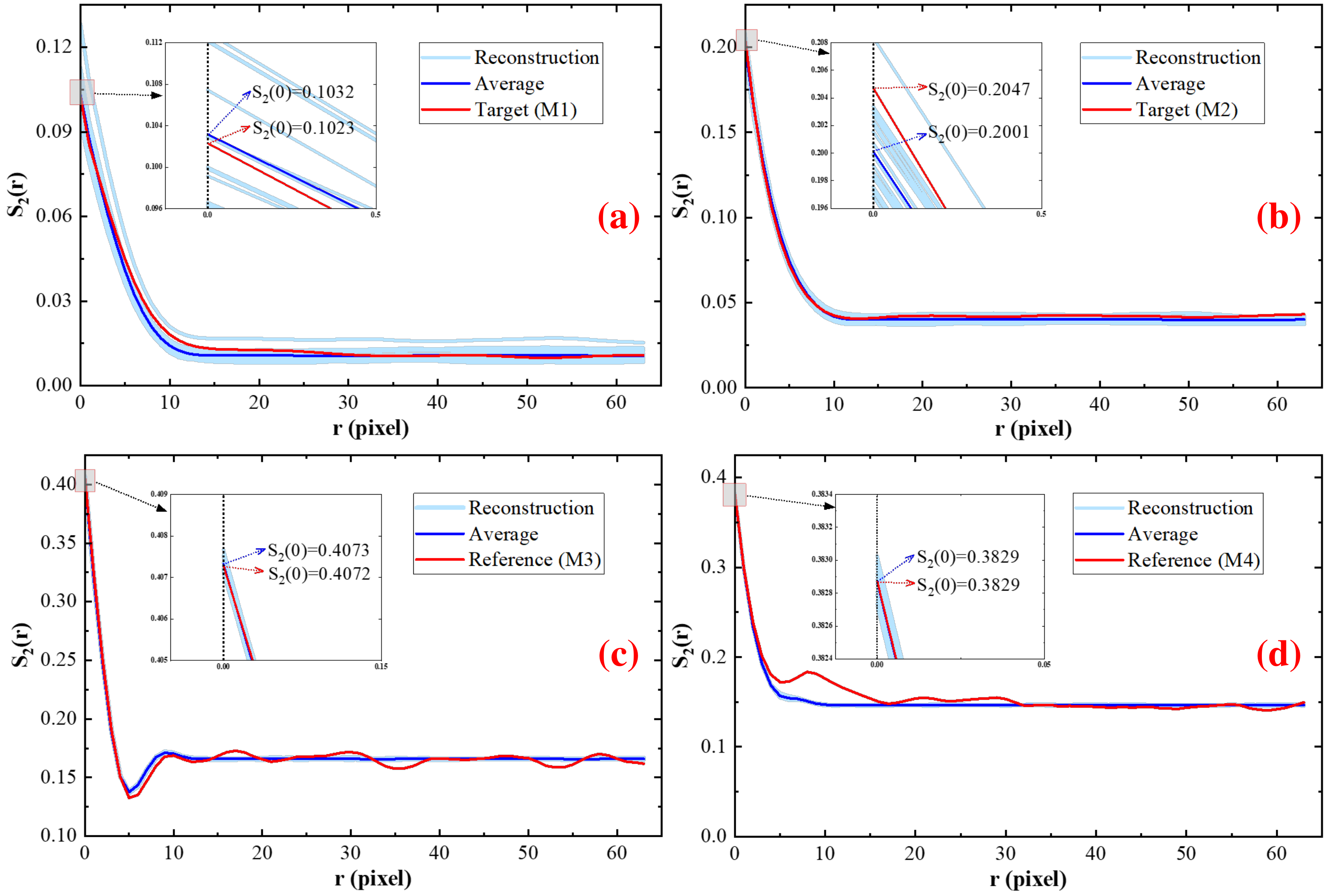}
	\caption{Comparisons of statistical correlation functions. (a) and (b) 2-point probability function curves between the reconstruction results and the target structure for silica material and porous rock; (c) and (d) 2-point probability function curves between the reconstruction results and the reference image for battery electrode material and superalloy. The subfigure is a zoom-in of the starting point region with annotations for the starting point value, phase volume fraction.}
	\label{fig:3-1-2}
\end{figure}

Figure \ref{fig:3-1-2}(a) and (b) compare the 2-point probability functions between the reconstruction results and the target structures for silica material and porous rock. The 2-point probability function curves for both the reconstruction samples and the target structures represent the average values across three directions X, Y, and Z. From Figure \ref{fig:3-1-2}(a) and (b), the statistical parameters of the reconstruction samples are closely aligned with those of the target structures. Despite some minor differences and inherent randomness, the overall shape and trend of the curves remain similar. The similarity in curve shapes and trends indicates that the spatial correlation of the reconstruction samples is close to that of the target structures. This suggests that our method accurately captures the essential microstructural characteristics.

Figure \ref{fig:3-1-2}(c) and (d) compare the 2-point probability functions between the reconstruction results and the reference images for battery electrode material and superalloy. The 2-point probability function curves for the reconstruction samples represent the average values across three directions (X, Y, and Z), while those for the reference images are averaged over two directions (X and Y). In the enlarged subfigures of Figure \ref{fig:3-1-2}(c) and (d), the statistical parameters of the reconstruction sample fluctuate within a small range around those of the reference images. The curve shape and trend for the battery electrode material are relatively close, indicating that the reconstruction sample effectively reproduces a similar spatial structure to the reference image. However, the 2-point probability function curves for superalloy show slight differences between the reconstruction samples and the reference image. This discrepancy is primarily due to the periodic and regular characteristics present in the reference image of superalloy, which pose a challenge for the model to fully reproduce using input noise structures with random initialization.

\begin{figure}%[htbp]
	\centering
        \includegraphics[scale=.15]{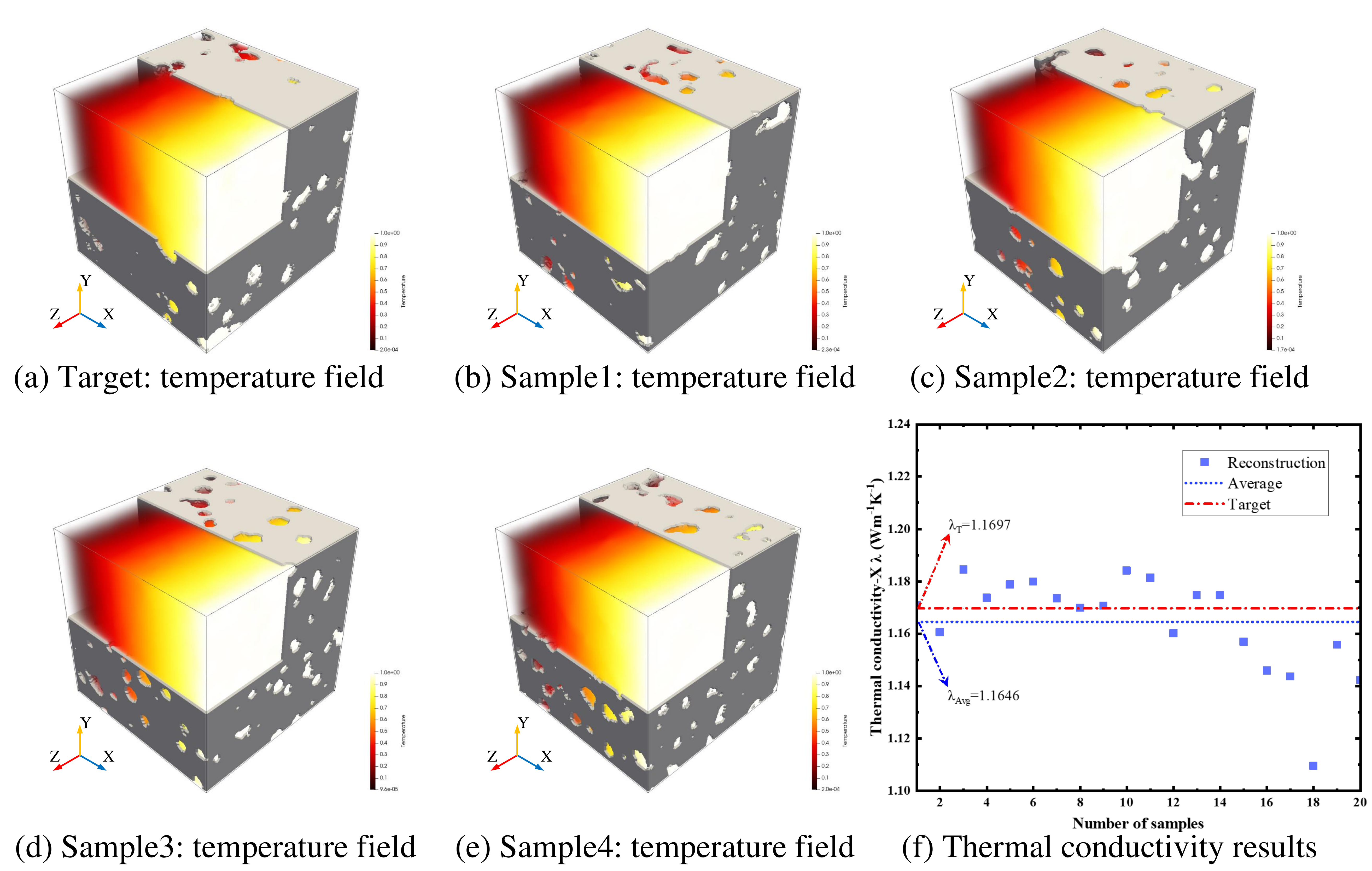}
	\caption{Thermal conductivity analysis of the target structure and the reconstruction results for silica material. (a) temperature field of the target structure, (b)–(e) temperature field of some reconstruction samples, and (f) comparison of thermal conductivity parameters of the target structure and the reconstruction samples.}
	\label{fig:3-1-3}
\end{figure}

To verify the consistency of physical properties between the reconstruction samples and the target structure, we employed the finite difference method (FDM) \citep{62_ferguson2018puma,63_ferguson2021update} to simulate the thermal conductivity of silica material. In the simulation setup, the thermal conductivity of the solid phase and pore phase and set to 1.4 and 0.026 W/m·K, and other parameters were set to default values. Additionally, periodic boundary conditions were applied for four side direction and constant value boundary condition were applied for the simulation direction \citep{64_hestenes1952methods}. The thermal conductivity was calculated for the target structure and 20 sets of the reconstruction results, and the temperature field was visualized for the target structure and partial reconstruction results, as shown in Figure \ref{fig:3-1-3}(a)–(e). The final thermal conductivity results are summarized and presented in Figure \ref{fig:3-1-3}(f).

As illustrated in Figure \ref{fig:3-1-3}(a)–(e), both the reconstruction samples and the target structure exhibit good consistency in the temperature field. They show similar downward trends and overall distributions in temperature, indicating that the reconstructed microstructures closely mimic the thermal behavior of the target structure. The average thermal conductivity parameters of the reconstruction samples, as shown in Figure \ref{fig:3-1-3}(f), are relatively close to those of the target structure. The values fluctuate around the thermal conductivity parameters of the target structure, further validating the consistency between the reconstruction results and the target microstructure.
\begin{figure}%[htbp]
	\centering
        \includegraphics[scale=.15]{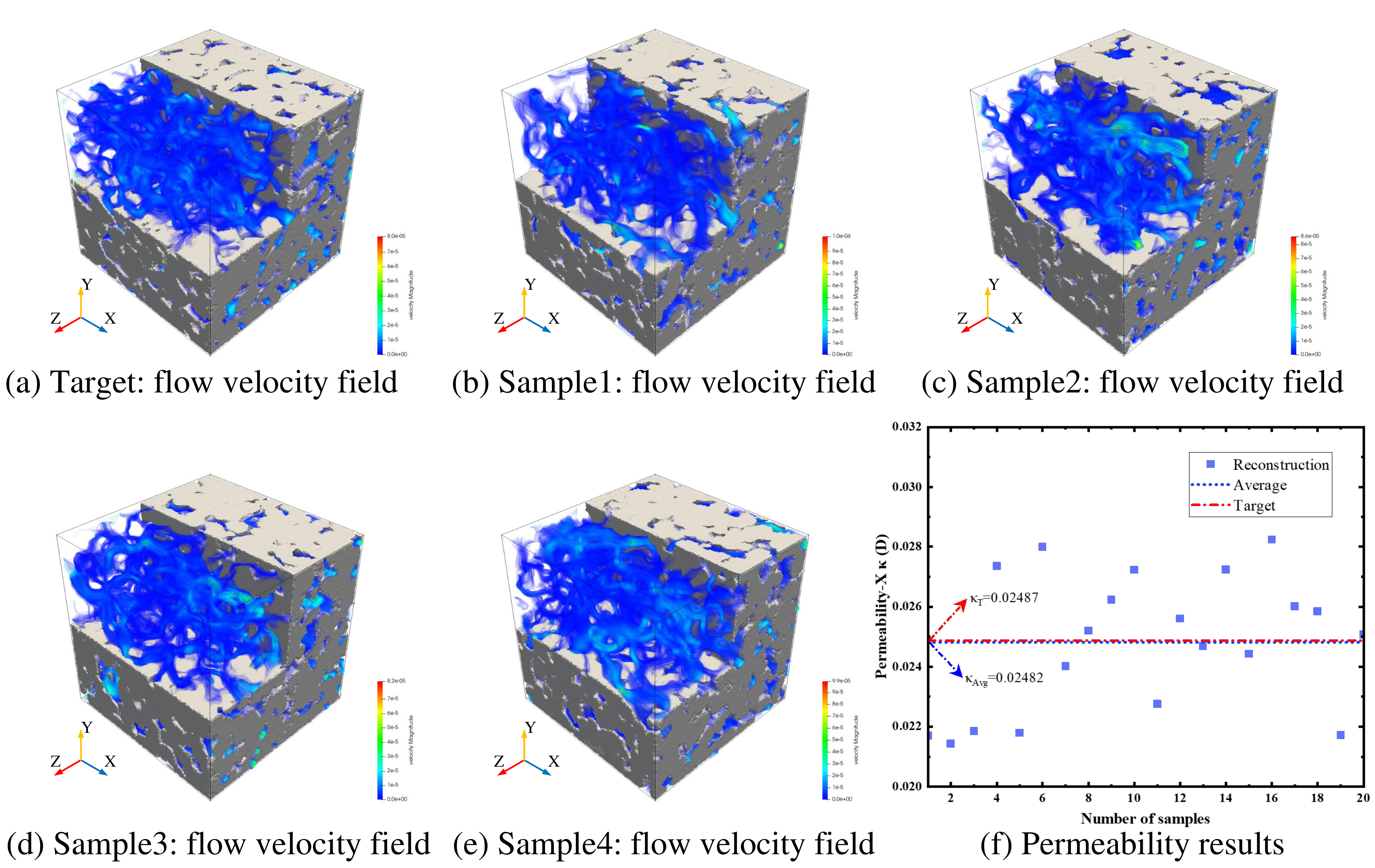}
	\caption{Permeability analysis of target structure and reconstruction results for porous rock. (a) flow velocity field of the target structure, (b)–(e) flow velocity field of some reconstruction samples, and (f) permeability parameters of the target structure and the reconstruction samples.}
	\label{fig:3-1-4}
\end{figure}

Similarly, to evaluate the consistency of permeability between the reconstruction results and the target structure, we utilized the lattice Boltzmann method (LBM) \citep{65_mcnamara1988use,66_santos2022mplbm} to simulate permeability in porous rock. In the simulation setup, the pressure is set to 0.005, other parameters were set to default values. Additionally, periodic boundary conditions \citep{67_latt2021palabos} were applied for the permeability calculation. The permeability was calculated for the target structure and 20 sets of reconstruction results, and the flow velocity field was visualized for the target structure and partial reconstruction results, as shown in Figure \ref{fig:3-1-4}(a)–(e). The final permeability results are summarized and presented in Figure \ref{fig:3-1-4}(f). As illustrated in Figure \ref{fig:3-1-4}(a)–(e), the reconstruction samples show good agreement with the target structure in terms of the flow velocity field. Both the velocity linear density and velocity distribution are closely matched, indicating that the reconstructed microstructures accurately capture the fluid dynamics of the target structure. Notably, the permeability parameter of the reconstruction samples fluctuates within a relatively large range around the permeability parameter of the target structure. This variability is primarily due to the complex local morphology and spatial structure of porous rock, which makes it challenging to fully reproduce the overall spatial connectivity. Despite this variability, the permeability of some reconstruction samples and the average permeability parameter (Figure \ref{fig:3-1-4}(f)) remain relatively close to the permeability of the target structure. This indicates that our method can effectively reproduce such complex spatial connectivity to a significant extent.

In summary, our method successfully completes model training and microstructure reconstruction using only a single reference image of the microstructure, without conditional parameters and control strategies. The reconstruction results exhibit excellent consistency with the target structure across visualization, statistical parameters, and physical properties aspects.

\subsection{Controllable reconstruction with conditional parameters}

To verify the ability of our method to generate microstructures that satisfy given conditional parameters, we conducted controllable reconstruction experiments incorporating conditional parameters and control strategies. Specifically, 2D-to-3D reconstructions with conditional parameters were performed on reference images of silica material, porous rock, battery electrode material, and superalloy. The parameter settings are identical to those used in the previous stochastic reconstruction experiments. Conditional parameters (target phase volume fractions) were introduced and set to 0.10, 0.15, 0.20, 0.25, and 0.30. For each group of reference images and each given conditional parameter, a separate model was trained using our method. In total, 20 models were trained, and each trained model was used to perform 5 reconstructions.

\begin{figure}%[htbp]
	\centering
        \includegraphics[scale=.15]{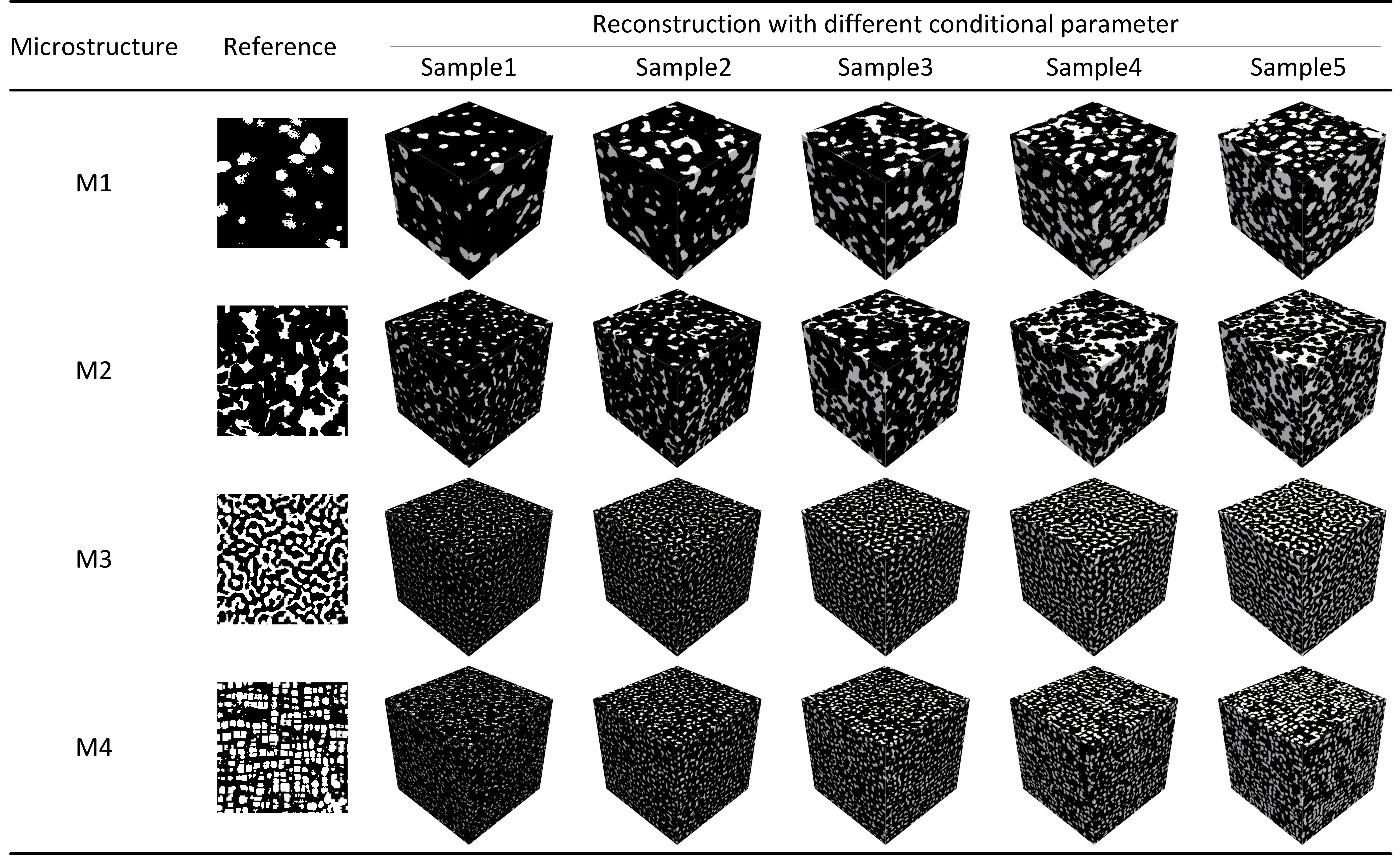}
	\caption{Visual comparison between the reference image and the reconstruction results of the controllable reconstruction with different conditional parameters. Microstructures M1–M4 are silica material, porous rock, battery electrode material, and superalloy, respectively. Samples 1–5 are the reconstruction samples under the conditional parameters of 0.10, 0.15, 0.20, 0.25 and 0.30, respectively.}
	\label{fig:3-2-1}
\end{figure}

Visual comparisons between the reference images and the reconstruction results in 2D-to-3D controllable reconstruction are shown in Figure \ref{fig:3-2-1}, where the microstructures M1–M4 are silica material, porous rock, battery electrode material, and superalloy, respectively. Both the reference images and the reconstruction samples are in a size of 128.

In the visual comparison presented in Figure \ref{fig:3-2-1}, our method successfully reproduces the local morphology of four different materials—silica material, porous rock, battery electrode material, and superalloy—even introducing conditional parameters. Importantly, our method can generate microstructures with different phase volume fractions by varying the conditional parameters.

\begin{figure}%[htbp]
	\centering
        \includegraphics[scale=.20]{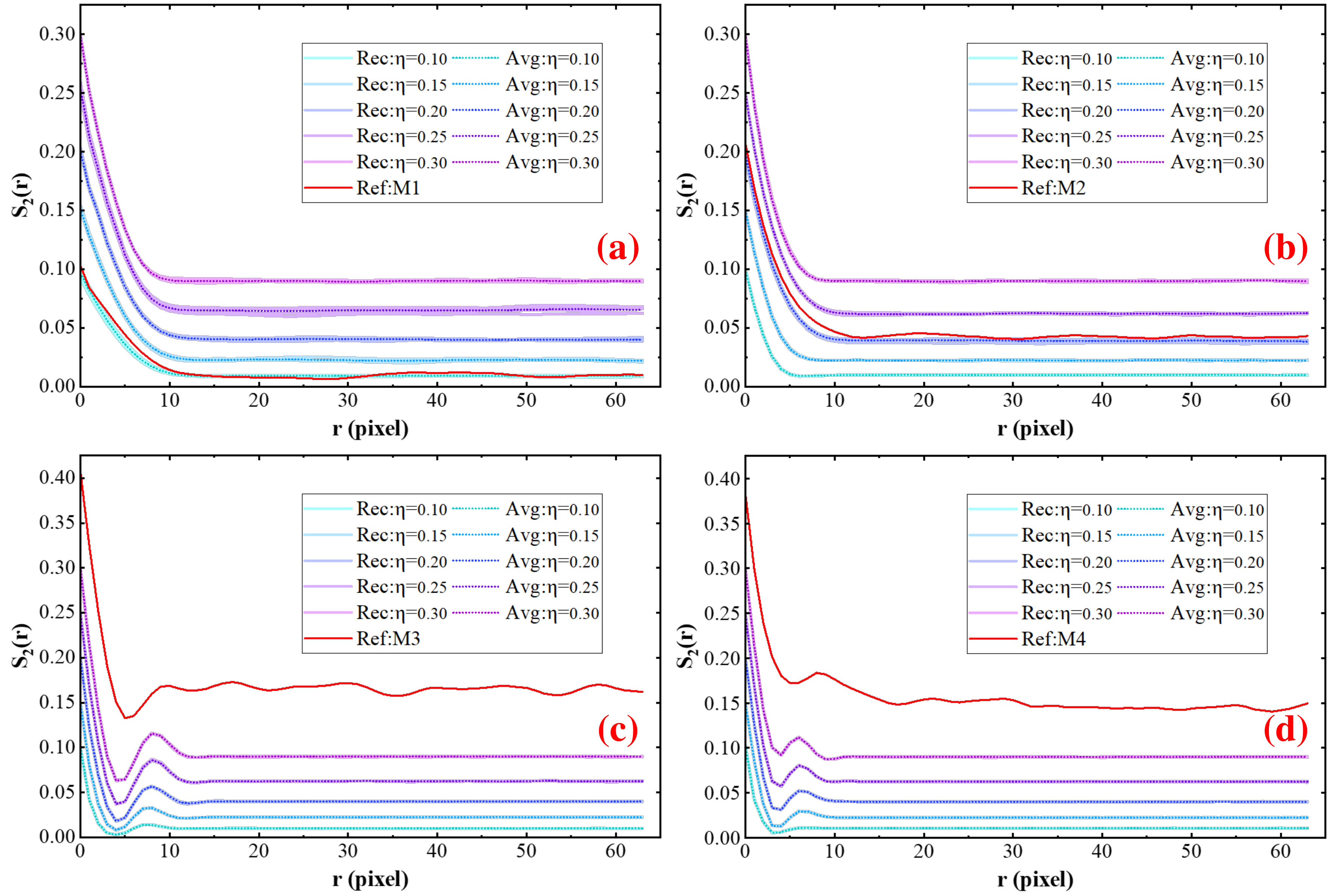}
	\caption{Comparisons of statistical correlation functions between the reconstruction results and the reference image under different conditional parameters. (a) 2-point probability function curve of silica material M1, (b) 2-point probability function curve of porous rock M2, (c) 2-point probability function curve of battery electrode material M3, and (d) 2-point probability function curve of superalloy M4.}
	\label{fig:3-2-2}
\end{figure}

To verify the consistency between the reconstruction results and the given condition parameters, we analyzed the statistical parameters of the reconstructed microstructures. Figure \ref{fig:3-2-2} presents the comparison of the 2-point probability functions between the reconstruction results and the reference images under different conditional parameters. Here, the two-point probability function curve for each reconstruction sample represents the average values across three directions (X, Y, and Z), while the two-point probability function curve for the reference images represents the average values over two directions (X and Y). 

As shown in Figure \ref{fig:3-2-2}(a)–(d), statistical parameters such as phase volume fractions (the starting point values) of the reconstruction samples are closely aligned with the given conditional parameters. The overall trend of the two-point probability function curves for the reconstruction samples is consistent with those of the reference images. This indicates that the reconstruction results maintain a similar spatial distribution to the reference images, even when different condition parameters are applied.

\begin{figure}%[htbp]
	\centering
        \includegraphics[scale=.15]{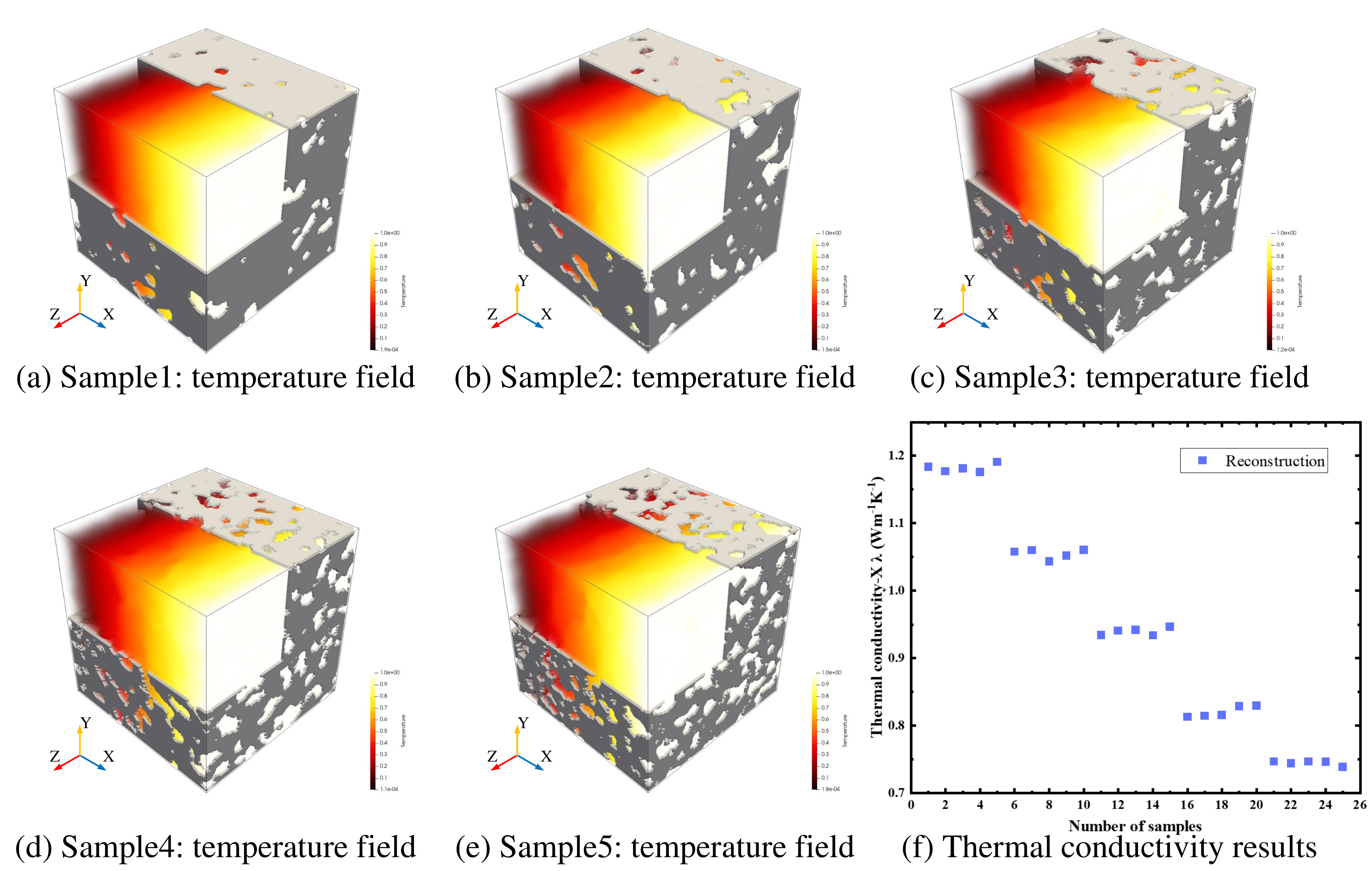}
	\caption{Thermal conductivity analysis of the reconstruction results with different condition parameters for silica material. (a)–(e) temperature field of reconstruction samples with conditional parameters of 0.10, 0.15, 0.20, 0.25, and 0.30, respectively, and (f) thermal conductivity results of 25 sets of reconstruction results.}
	\label{fig:3-2-3}
\end{figure}

Finally, to evaluate the physical properties of the reconstruction results under different conditional parameters, we simulated the thermal conductivity of silica material using the FDM. The parameter settings for these simulations were identical to those used in the previous stochastic reconstructions without conditional parameters. In this simulation, the thermal conductivity was calculated for 25 sets of reconstruction results, and the temperature field of partial reconstruction results is visualized in Figure \ref{fig:3-2-3}(a)–(e), each corresponding to different conditional parameters.

As shown in Figure \ref{fig:3-2-3}(a)–(e), reconstruction samples with different conditional parameters exhibit different downward trend in temperature field distributions, reflecting variations in microstructural characteristics introduced by the conditional parameters. The final thermal conductivity results are summarized and presented in Figure \ref{fig:3-2-3}(f). For each set of conditional parameters, the thermal conductivity parameters of the 5 reconstruction samples are relatively consistent, indicating that our method can generate microstructures with controlled thermal properties.

\begin{figure}%[htbp]
	\centering
        \includegraphics[scale=.15]{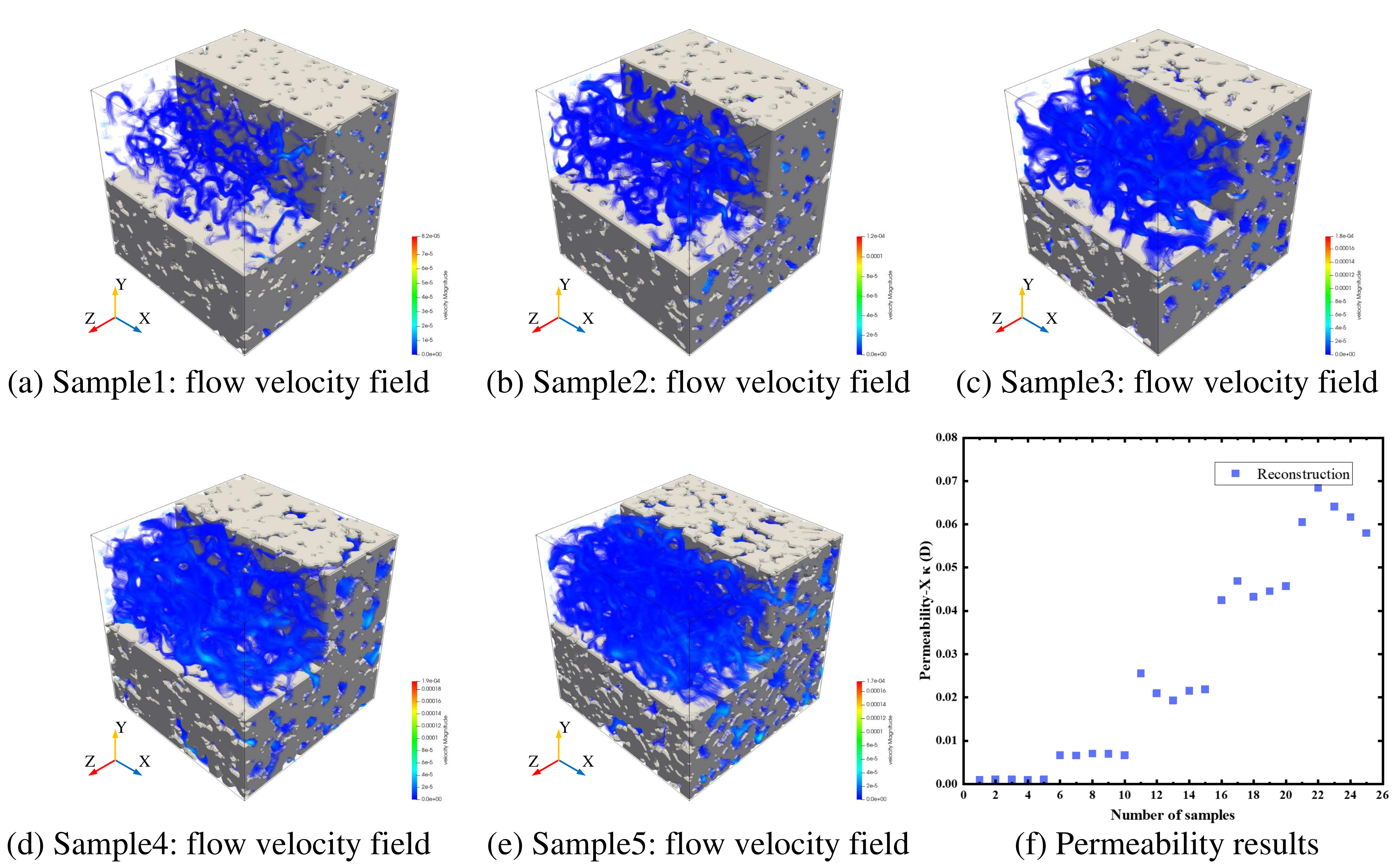}
	\caption{Permeability analysis of the reconstruction results with different conditional parameters for porous rock. (a)–(e) flow velocity field of reconstruction samples with conditional parameters of 0.10, 0.15, 0.20, 0.25, and 0.30, respectively, and (f) permeability results of 25 sets of reconstruction results.}
	\label{fig:3-2-4}
\end{figure}

Similarly, we used the LBM to simulate the permeability of reconstruction results for porous rock under different conditional parameters. The parameter settings were identical to those used in the previous stochastic reconstructions without conditional parameters. In this simulation, the permeability was calculated for 25 sets of reconstruction results, and the flow velocity field of partial reconstruction results is visualized in Figure \ref{fig:3-2-4}(a)–(e), each corresponding to different conditional parameters.

As shown in Figure \ref{fig:3-2-4}(a)–(e), Samples 1–5 exhibit distinct flow velocity field distributions under different conditional parameters. Specifically, the larger the given conditional parameter (pore phase volume fraction), the more and denser the streamlines of the flow velocity field become. The final permeability parameter results are summarized and presented in Figure \ref{fig:3-2-4}(f). For each set of conditional parameters, the permeability results of the 5 reconstruction samples exhibit some variability within a slightly larger range but are still relatively consistent.

In summary, by introducing conditional parameters and control strategies, our method achieves controllable characterization and reconstruction from a single microstructure image. This approach enables the generation of microstructures that not only exhibit similar local morphology to the reference images but also satisfy the specified conditional parameters. By controlling these parameters, we can ultimately produce microstructures with distinct physical properties, such as thermal conductivity and permeability.

\subsection{Reconstruction of spatially heterogeneous microstructures}

One advantage of using 3D structure as input for microstructure reconstruction is the ease with which its spatial distribution can be controlled. To accomplish spatially heterogeneous microstructure reconstruction tasks, we introduce an additional spatial location mask to control the spatial distribution of the initialization input. Mathematically, this process can be expressed as $Y_{Ms} = G(X \cdot M_s; \theta)$, where $M_s$ is the spatial location mask, $X$ is the input structure, $Y_{Ms}$ is the generated microstructure, and $G$ is the trained neural network model parameterized by $\theta$.

By incorporating spatial location masks, our method can generate microstructures that closely match the specified spatial distribution. This allows for precise control over the spatial heterogeneity of the reconstructed microstructures. The spatial location masks can be user-defined, derived from existing material microstructures, or generated based on statistical features extracted from material images. This flexibility enables users to tailor the spatial distribution according to specific requirements or replicate complex patterns in real materials.

\begin{figure}%[htbp]
	\centering
        \includegraphics[scale=.15]{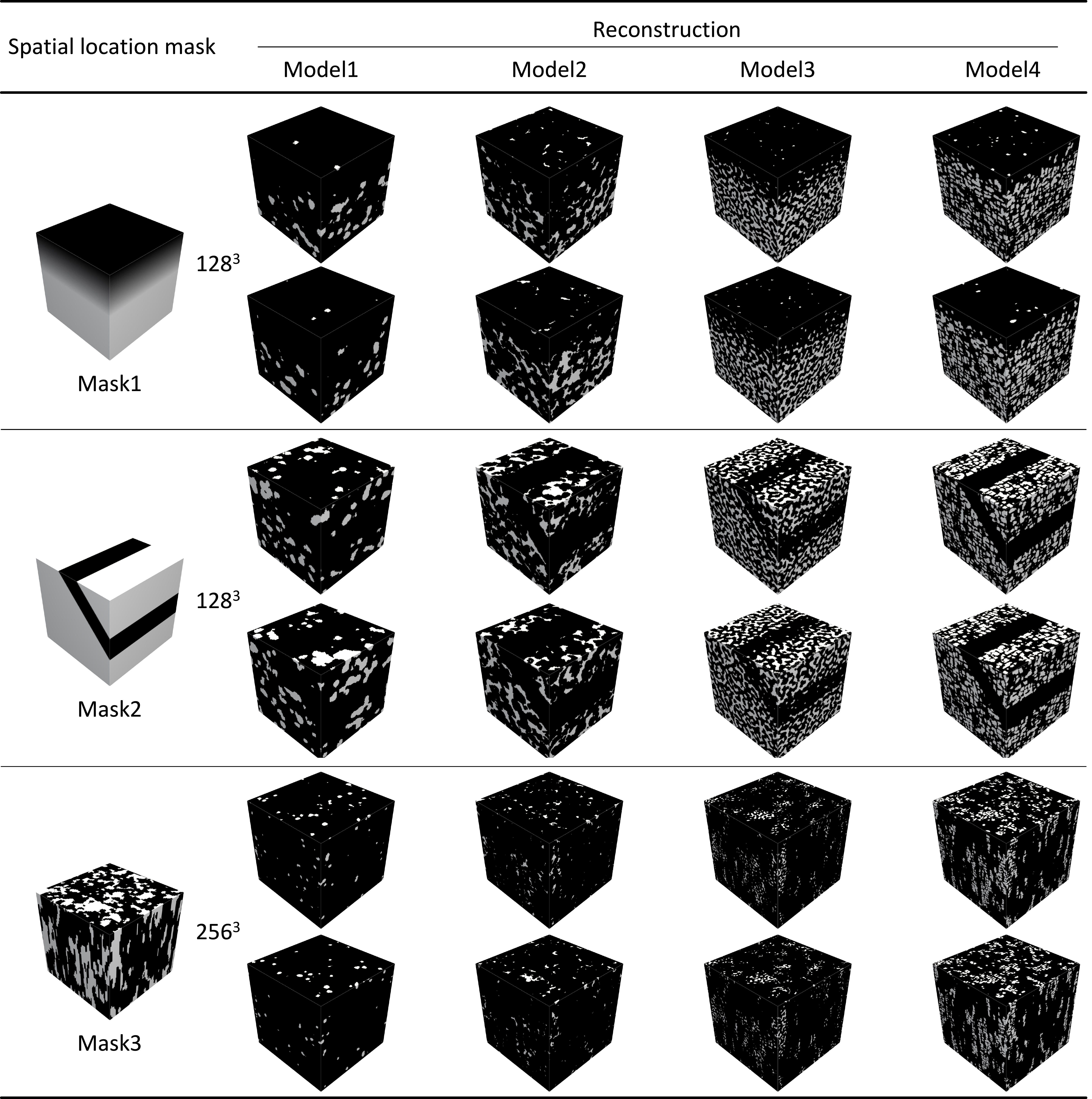}
	\caption{Visualization of the reconstruction results of spatially heterogeneous microstructures. Mask 1 and 2 are user-defined; Mask 3 is the microstructure of synthetic ceramic; Models 1–4 are the models trained from reference images of silica material, porous rock, battery electrode material, and superalloy, respectively.}
	\label{fig:3-3-1}
\end{figure}

Specifically, three types of spatial location masks were utilized in the reconstruction experiments of spatially heterogeneous microstructures. As shown in Figure \ref{fig:3-3-1}, the first two mask types are user-defined, and the third mask type employs the microstructure of synthetic ceramic material \citep{59_bostanabad2016characterization}, offering a practical example of spatial heterogeneity. The reconstruction experiments were conducted using models previously trained on reference images of four materials (silica material, porous rock, battery electrode material, and superalloy).

In Figure \ref{fig:3-3-1}, the reconstructed microstructures maintain the spatial distribution dictated by the original spatial location masks, and they exhibit local morphological details that closely resemble those of the corresponding reference images. These findings indicate that our method can effectively generate special spatially heterogeneous microstructures by introducing spatial location masks. Furthermore, by using existing microstructures or those generated by stochastic reconstruction methods as spatial location masks, our method has the potential to reconstruct even more complex microstructures. This process shares similarities with local–global decoupling reconstruction methods \citep{37_robertson2023local,68_murgas2024modeling}. The overall spatial structure of the complex microstructure is first established and introduced as a spatial location mask, and the local morphological details of the complex microstructure are subsequent reconstructed.

\begin{figure}%[htbp]
	\centering
        \includegraphics[scale=.34]{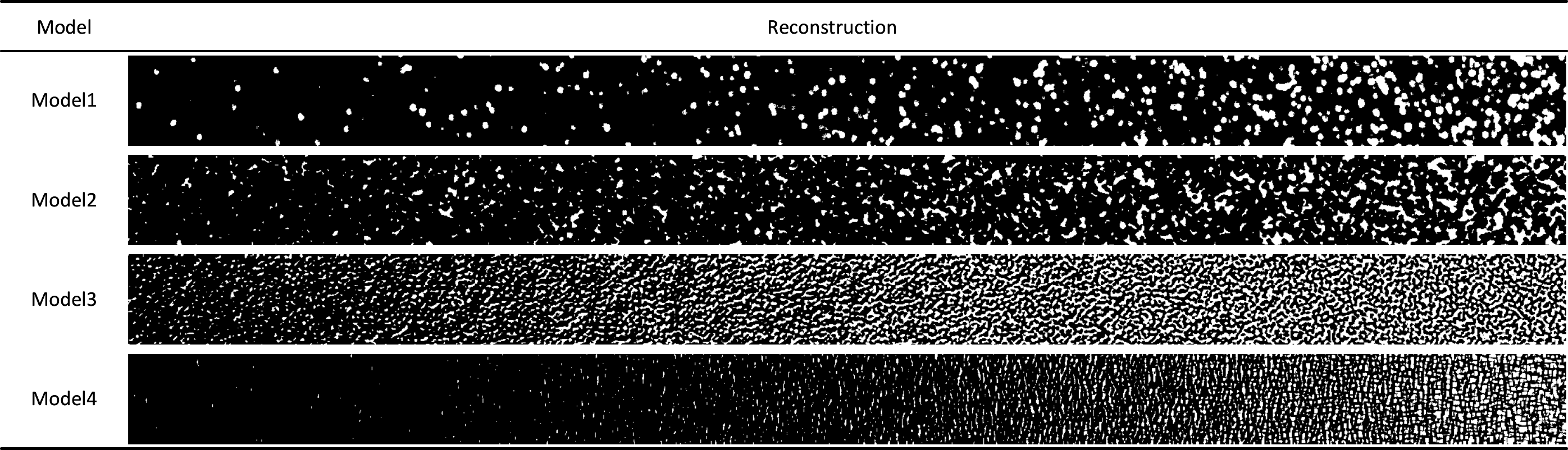}
	\caption{Visualization of the reconstruction results from continuous sampling of Gaussian distributions with different means. Models 1–4 are the models trained from reference images of silica material, porous rock, battery electrode material, and superalloy, respectively.}
	\label{fig:3-3-2}
\end{figure}

In addition, our method achieves controllable generation by adjusting the distribution of the initialized input structures, without explicitly obtaining the latent vector of the microstructure feature space. Figure \ref{fig:3-3-2} illustrates the reconstruction results obtained by feeding structures continuously sampled from Gaussian distributions with different means into the trained model. This approach allows for continuous and diverse microstructure generation. On the one hand, sampling from Gaussian distributions with varying means and variances enables the generation of microstructures corresponding to different statistical distributions. On the other hand, by interpolating the initialization distribution, it is possible to generate microstructures that exhibit continuous transitions.

\subsection{Large-size microstructure reconstruction}

The overall network architecture of our method primarily employs convolutional layers and does not include any fully connected layers. This design enables the network to adapt to inputs of various sizes. However, there are limitations when dealing with extremely large 3D microstructures, such as those with sizes of 512 or 1024, due to constraints in computer memory resources. Notably, our network architecture does not incorporate boundary filling conditions. Without built-in boundary conditions, reconstructing microstructures with periodic boundaries becomes challenging (generating these microstructures requires the implementation of additional periodic boundary conditions, either within the convolutional layers or applied to the input structures). Despite this limitation, the network can handle large-size microstructure reconstruction tasks using a chunking strategy. By dividing the large microstructure into multiple smaller blocks, the network can reconstruct each block individually. This strategy circumvents memory constraints and allows for the reconstruction of large-size microstructures while maintaining the integrity of the overall structure.

Specifically, in the chunking strategy, our method sets the reconstruction block size to 128. Consequently, a 512-size microstructure can be divided into 64 sub-blocks, while a 1024-size microstructure can be divided into 512 sub-blocks for reconstruction, respectively. The reconstruction process utilizes the trained model from the previous stochastic reconstruction experiment. Visualizations of the 1024-size reconstruction results are presented in Figure \ref{fig:A2-1}–\ref{fig:A2-4} in the Appendix.

All the above experiments were conducted on a computer equipped with an NVIDIA GeForce RTX 4070 Ti graphics card. Table \ref{tab:1} summarizes the model training times and the average time required for the model to perform a single reconstruction in both the stochastic reconstruction experiment and the large-size reconstruction experiment. Due to the utilization of pre-trained neural network models, the entire reconstruction process, even for large-size microstructures, primarily involves multiple forward mappings, resulting in relatively high reconstruction efficiency.

Finally, given a microstructure reference image and assuming $k$ instances of $N$-size microstructure reconstruction, the overall reconstruction efficiency $\tau$ of our method can be approximately estimated using Eq. \ref{Eq:22}. 
\begin{equation}\label{Eq:22}
\tau  = {\tau _{train}} + k \times {\left( {{N \mathord{\left/
 {\vphantom {N {128}}} \right.
 \kern-\nulldelimiterspace} {128}}} \right)^3} \times {\tau _{rec}}
\end{equation}
where $\tau _{train}$ represents the time required to train the model, and $\tau _{rec}$ denotes the average reconstruction time for 128-size microstructures using the trained model. As the reconstruction size increases, the overall reconstruction efficiency $\tau$ will exhibit an approximate polynomial increase.
\begin{table}%[h]
\centering
\caption{Training and reconstruction time analysis for different microstructures}
\begin{tabular}{@{}ccccc@{}}
\toprule
\multirow{2}{*}{Microstructure} & \multirow{2}{*}{Training time(s)} & \multicolumn{3}{c}{Reconstruction time(s)} \\ \cmidrule(l){3-5}
               &                  & 128-size & 512-size & 1024-size \\ \midrule
M1             & 200.93           & 0.94     & 16.11    & 139.79     \\
M2             & 200.79           & 0.94     & 17.25    & 143.66     \\
M3             & 197.95           & 0.90     & 17.76    & 156.53     \\
M4             & 200.56           & 0.92     & 17.87    & 147.94     \\ \bottomrule
\end{tabular}
\label{tab:1}
\end{table}

\section{Discussion}\label{sec4}

Accuracy, controllability, generalization, and heterogeneous reconstruction are several key points in microstructure characterization and reconstruction. In these aspects, our method exhibits some advantages while also facing certain limitations.

\textbf{Accuracy.} Our method adopts local pattern distribution to characterize the microstructure, rather than relying on statistical correlation functions or feature extraction networks. Compared to statistical correlation functions such as autocorrelation functions \citep{57_ma2023fast}, local pattern distribution can more accurately describe the local morphology of microstructures. Moreover, compared to feature extraction networks, this descriptor ensures similar accuracy to the VGG network \citep{69_DBLP:journals/corr/SimonyanZ14a} but offers advantages in characterizing and reconstructing periodic microstructures, along with stronger interpretability. Figure \ref{fig:A3} in the Appendix presents a comparison of the training process for some regular periodic structures under the same network structure and parameter settings, which contrasts the loss functions of autocorrelation function-based loss, VGG network-based loss, and local pattern distribution-based loss.

\textbf{Controllability and generalization.} Our method employs a control strategy to sample a new target distribution that satisfies given conditional parameters from the original local pattern distribution. This target distribution is then used as the optimization objective for model training and reconstruction. While this approach achieves some degree of controllability and generalization, it has limitations when dealing with multiple conditional parameters and corresponding multiple target distributions. Specifically, when multiple conditional parameters are provided, resulting in multiple target distributions generated according to the control strategy, a single neural network struggles to learn the mappings between multiple input distributions and multiple target distributions simultaneously. This difficulty arises because it is challenging to independently express and decouple these objective distributions within a single model. Consequently, achieving simultaneous minimization of multiple different optimization objectives becomes problematic.

Figure \ref{fig:A4} in the Appendix illustrates this issue by comparing the loss iteration processes of simultaneously fitting multiple different Gaussian distributions and multiple different target distributions and separately fitting each Gaussian distribution and each target distribution individually. When attempting to fit multiple mappings simultaneously, the neural network achieves a certain level of reconstruction performance but falls short of the accuracy attainable with single input-to-target distribution mappings. Figure \ref{fig:A5} in the Appendix provides visualizations of sections in one direction during the iteration process of fitting multiple mappings directly versus fitting a single mapping separately. To address these limitations, decoupling the target distributions and employing advanced training strategies and techniques will be crucial. Enhancing the ability of a single neural network model to generalize across multiple input and target distributions will be a key focus in subsequent work.

\begin{figure}%[htbp]
	\centering
        \includegraphics[scale=.32]{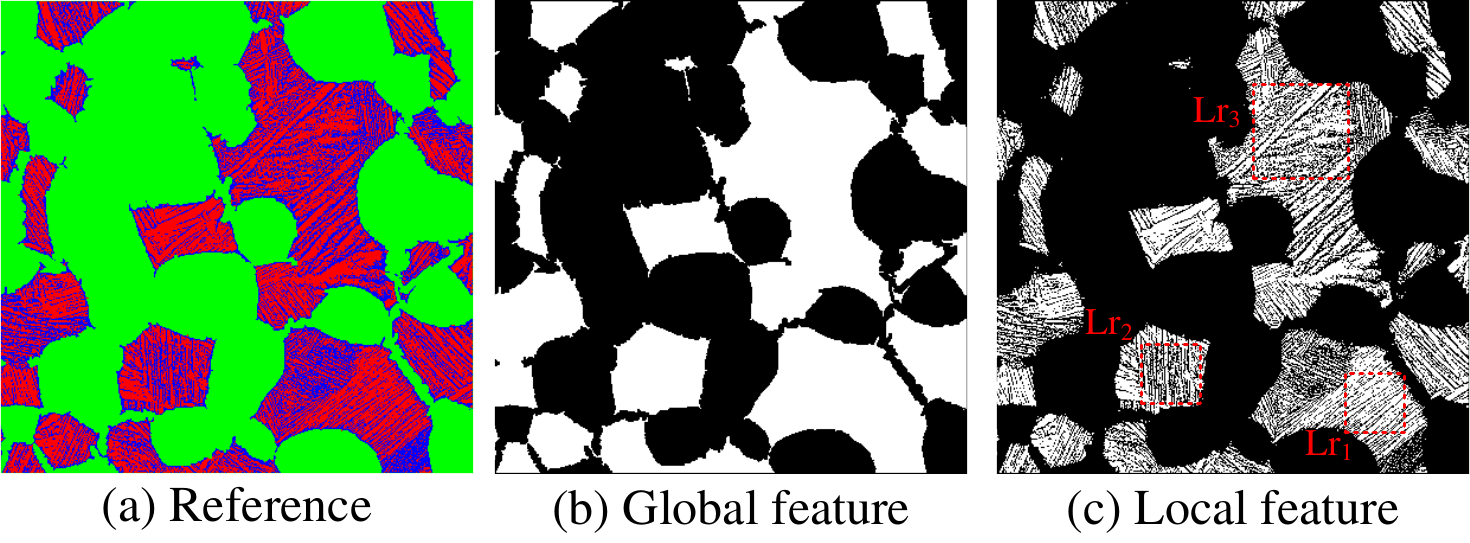}
	\caption{Reference image and local–global decoupling features of three-phase Titanium microstructure. (a) Reference image, (b) Global feature image, and (c) Local feature image. Lr1, Lr2, and Lr3 are three local regions used to extract local pattern distributions.}
	\label{fig:4-1-1}
\end{figure}

\begin{figure}%[htbp]
	\centering
        \includegraphics[scale=.5]{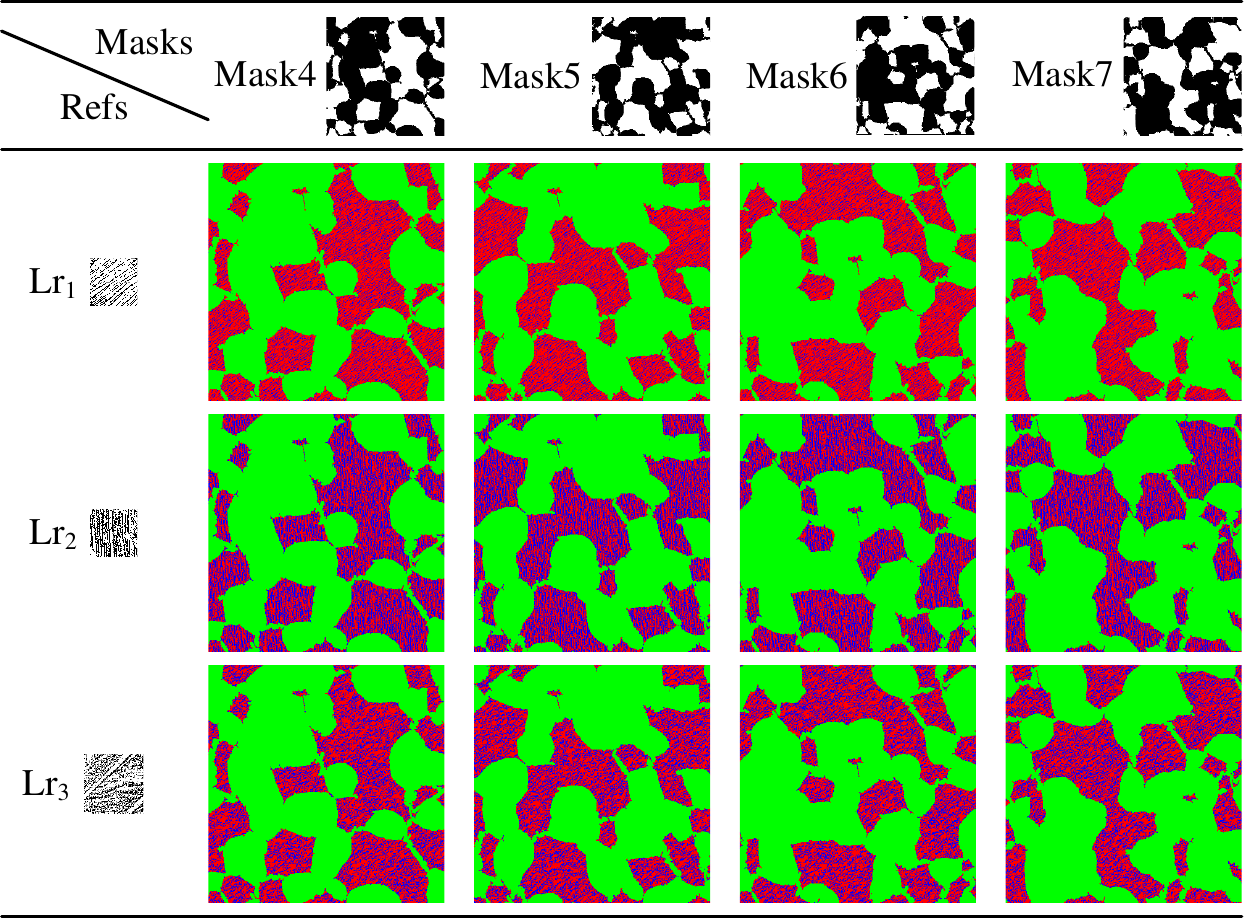}
	\caption{Visualization of reconstruction results for three-phase Titanium microstructure. Lr1, Lr2, and Lr3 are three local regions used to train neural network models separately. Mask4 is global feature image of Titanium microstructure, and Mask5–7 are randomly generated statistical equivalent microstructures.}
	\label{fig:4-1-2}
\end{figure}

\textbf{Heterogeneous and complex microstructure reconstruction.} Our method introduces spatial location masks to generate spatially heterogeneous microstructures and serve complex microstructure generation with local–global decoupling. Here, a simple case study involving three-phase Titanium microstructure reconstruction \citep{70_millan2022study,37_robertson2023local} illustrates this potential application. Firstly, the global feature and local feature of the three-phase Titanium microstructure (Figure \ref{fig:4-1-1}(a)) are extracted. Notably, the green phase part corresponds precisely to the global feature (Figure \ref{fig:4-1-1}(b)), while the red phase part represents the local feature (Figure \ref{fig:4-1-1}(c)). Secondly, regions representing local features are extracted to obtain the corresponding local pattern distribution as the optimization target. Here, three different local regions—Lr1, Lr2, and Lr3—are chosen from Figure \ref{fig:4-1-1}(c) to train the neural networks separately. Subsequently, the global feature is used as the spatial position mask (Mask4 in Figure \ref{fig:4-1-2}). Alternatively, statistically equivalent microstructures are randomly generated as spatial position masks (Mask5–7 in Figure \ref{fig:4-1-2}). Finally, the trained neural network model incorporates the spatial position mask for reconstruction. In the final three-phase reconstruction results, the green phase part is determined by the spatial location mask, the red phase part is generated by the trained neural network, and the blue phase part represents the remaining region.

The final reconstruction results of the three-phase Titanium microstructure are shown in Figure \ref{fig:4-1-2}. While our method only learns the local pattern distribution at the center region, leading to less accurate reproduction of local patterns at the edge regions, it still achieves a satisfactory overall reconstruction effect. Specifically, it successfully reproduces the global structure and local details similar to the target structure for complex microstructure generation tasks. A critical issue in practice is the acquisition of accurate spatial location masks for complex microstructures, which requires precise local feature region segmentation and extraction methods. Therefore, combining suitable local feature region segmentation and extraction techniques will be key to improving the performance of heterogeneous and complex microstructure reconstruction in future studies.

\section{Conclusion}\label{sec5}

In this paper, we propose a statistically controllable microstructure reconstruction framework that combines the neural networks with sliced-Wasserstein metric. Specifically, our approach utilizes local pattern distribution for microstructure characterization and employs a controlled sampling strategy to generate a new target distribution that satisfies given conditional parameters from the original local pattern distribution. A neural network-based model establishes the mapping between the input distribution and the target local pattern distribution, enabling microstructure reconstruction. To optimize the reconstruction process, we use sliced-Wasserstein metric and gradient optimization techniques to minimize the distance between the local pattern distribution of the reconstruction results and the target distribution. This iterative process updates the model parameters, leading to a stable and reliable reconstruction model.

Our method offers several key advantages. (1) It can complete microstructure stochastic reconstruction tasks even in cases of small sample sizes. (2) It generates microstructures that satisfy specified conditional parameters for controllable reconstruction. (3) By introducing spatial location masks, our method excels at generating spatially heterogeneous and complex microstructures. Multiple reconstruction experiments have demonstrated the effectiveness of our method. These findings hold significant promise for advancing structure–property linkage studies and computer-aided design.

While our method shows promising results, future research will focus on enhancing its capabilities by addressing current limitations, such as improving edge region reproduction and developing more sophisticated spatial location mask generation techniques. Additionally, exploring advanced training strategies and multi-task learning approaches could further improve the generalization and accuracy of our model.

\section{Acknowledgments}

The authors would like to acknowledge the support of the National Natural Science Foundation of China (Grant No. 62071315) and Sichuan Science and Technology Program (Grant No.2024ZYD0263). 

\newpage

\textbf{Code availability section}

Name of the code: Statistically controllable microstructure reconstruction framework.

Contact: hxh@scu.edu.cn

Program language: Python.
 
Software required: Python 3.10+, PyTorch (with CUDA support recommended), TorchVision, OpenCV, NumPy, Pillow.

The source codes are available for downloading at the link: https://github.com/scu-stu-mzc/Statistically-controllable-microstructure-reconstruction-framework.

\section*{Appendix}\label{secA1}
%\appendix
\setcounter{figure}{0}
\renewcommand{\thefigure}{A.\arabic{figure}}
%\section*{Appendix}

\textbf{A.1 Comparisons of statistical correlation functions for stochastic reconstruction}
\begin{figure}%[htbp]
	\centering
        \includegraphics[scale=.25]{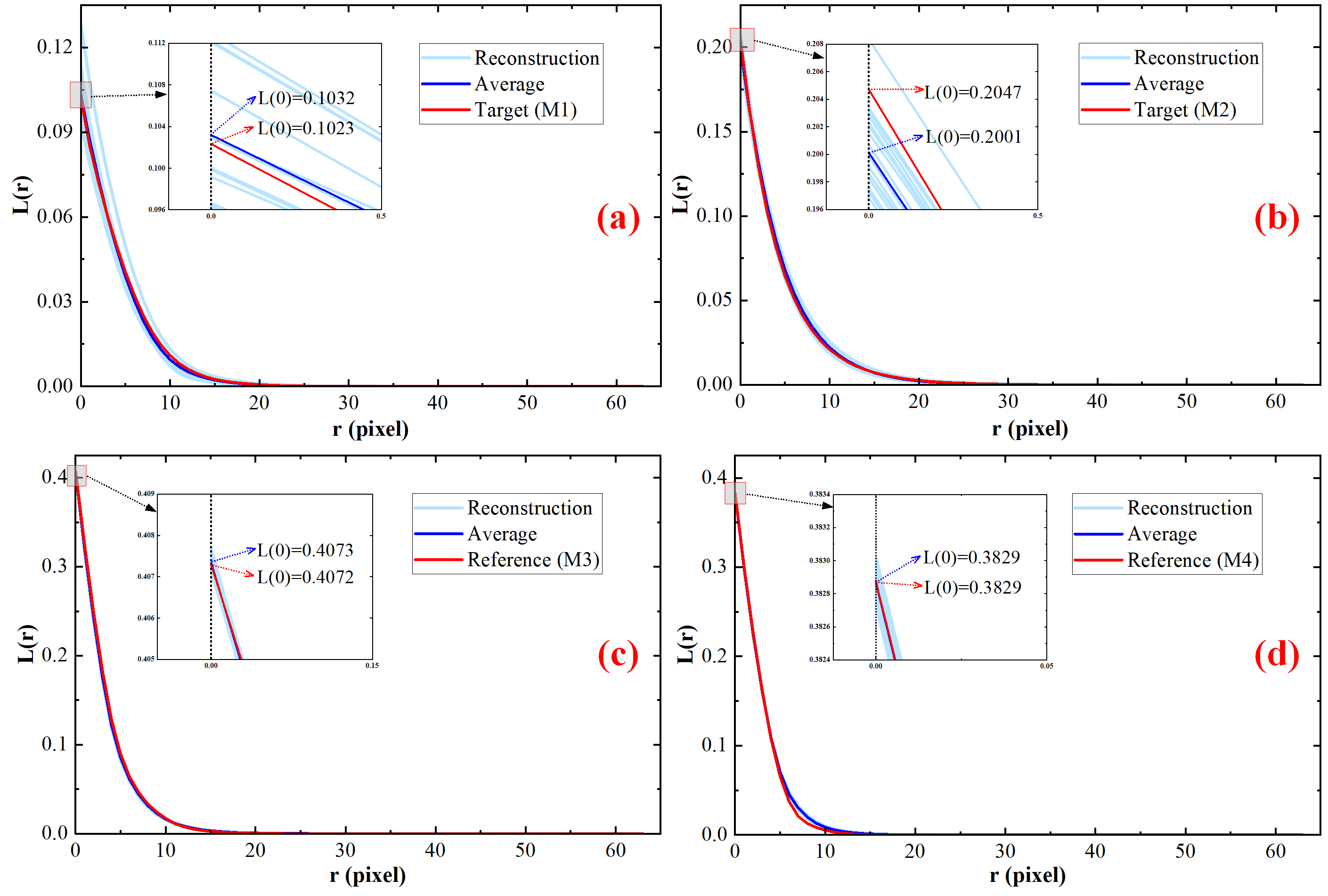}
	\caption{Comparisons of statistical correlation functions. (a) and (b) linear path function curves between the reconstruction results and the target structure for silica material and porous rock; (c) and (d) linear path function curves between the reconstruction results and the reference image for battery electrode material and metal material. The subfigure is a zoom-in of the starting point region with annotations for the starting point value, phase volume fraction.}
	\label{fig:A1}
\end{figure}

\newpage
\textbf{A.2 Visualization of large-size reconstruction results}
\begin{figure}%[htbp]
	\centering
        \includegraphics[scale=.25]{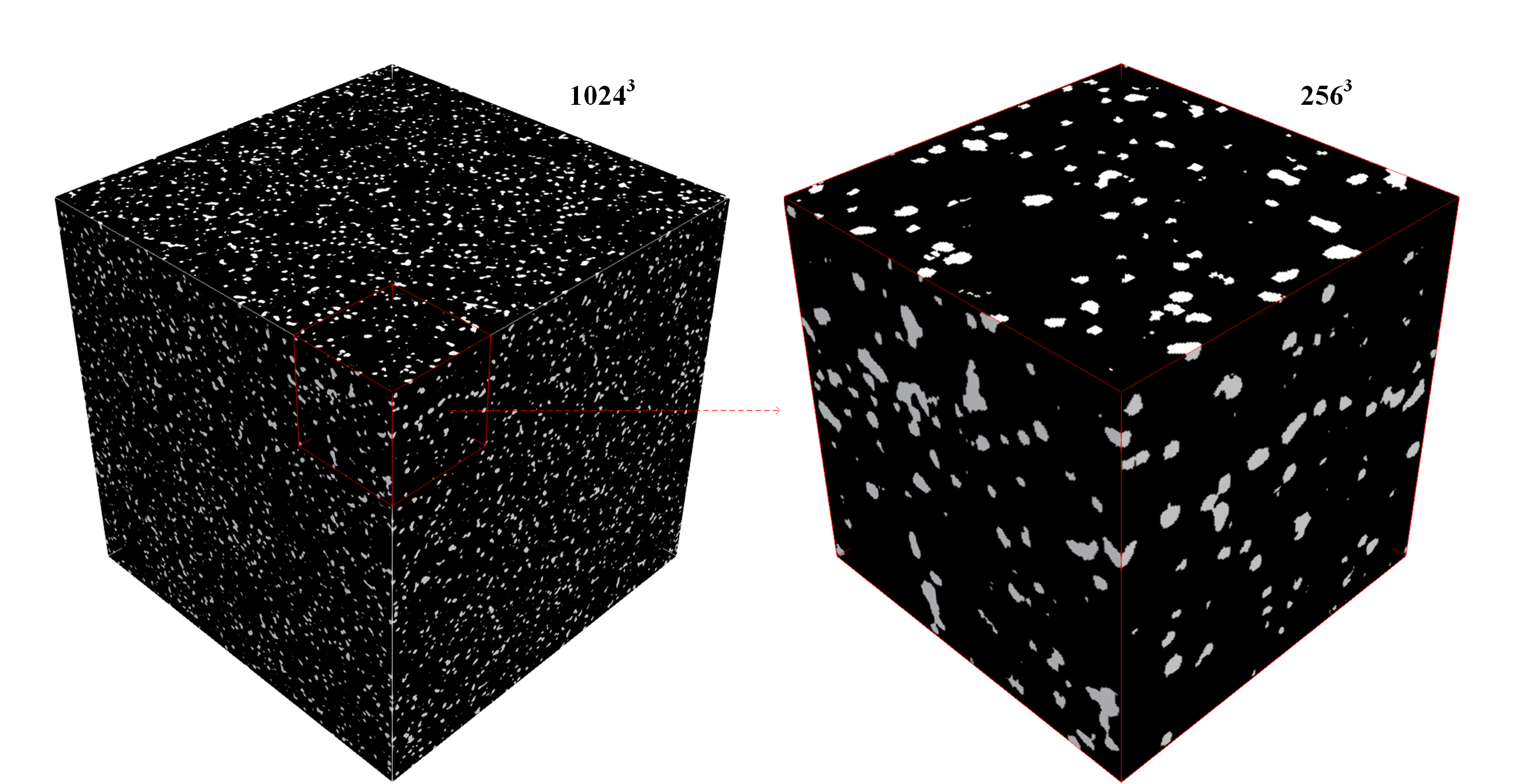}
	\caption{Visualization of a 1024-size reconstruction result and a cropped 256-size sub-block for silica material.}
	\label{fig:A2-1}
\end{figure}
\begin{figure}%[htbp]
	\centering
        \includegraphics[scale=.25]{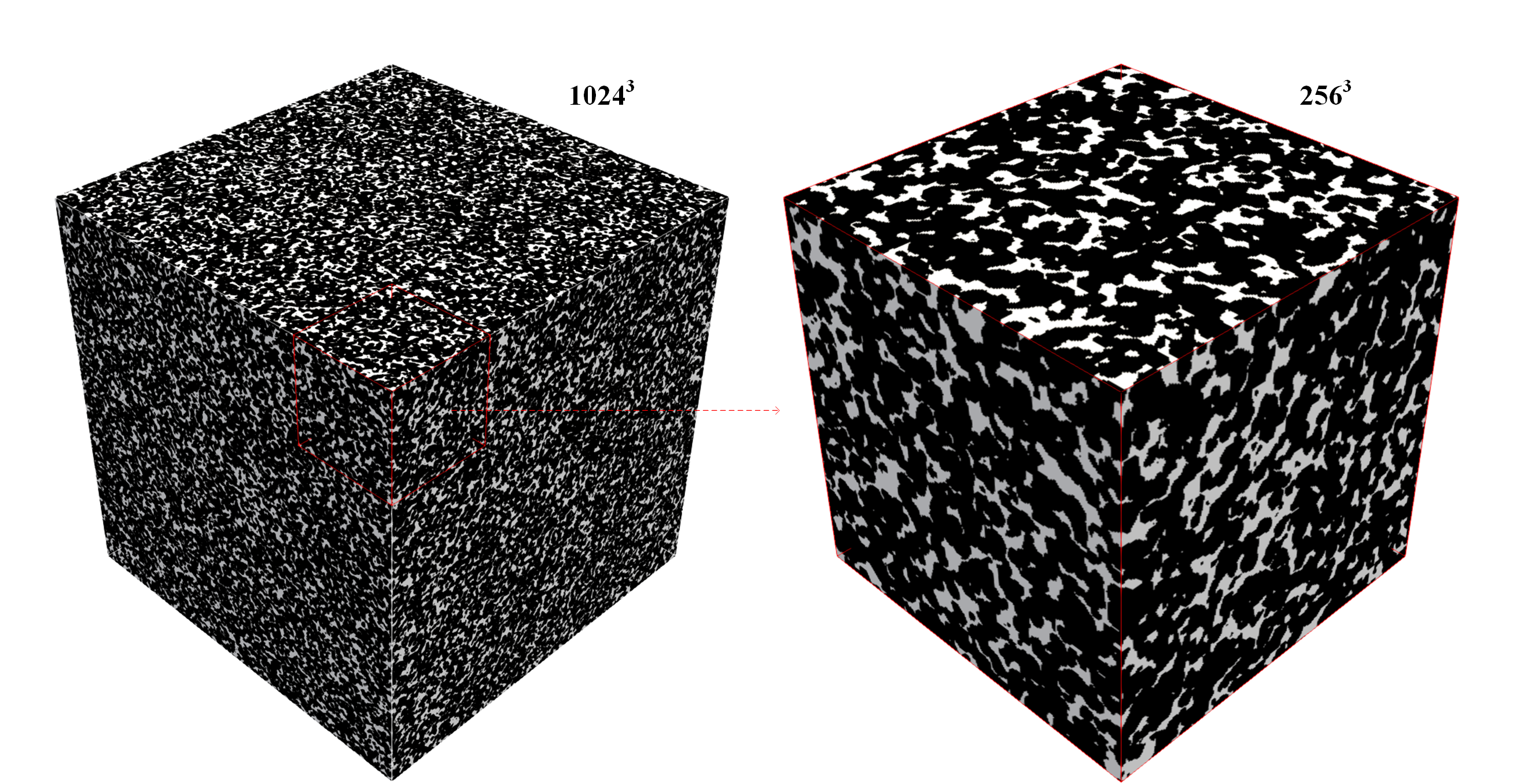}
	\caption{Visualization of a 1024-size reconstruction result and a cropped 256-size sub-block for porous rock.}
	\label{fig:A2-2}
\end{figure}

\newpage
\begin{figure}%[htbp]
	\centering
        \includegraphics[scale=.25]{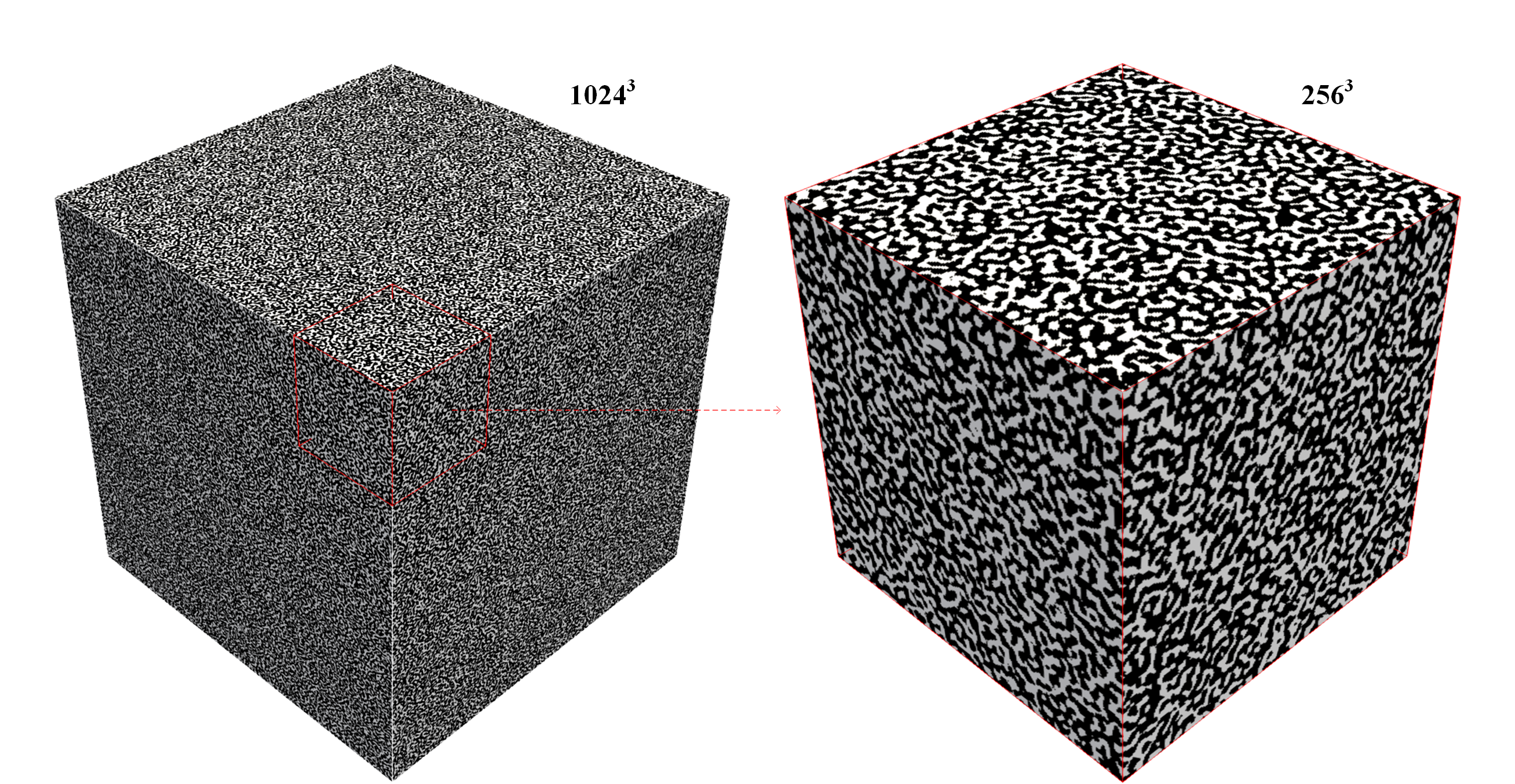}
	\caption{Visualization of a 1024-size reconstruction result and a cropped 256-size sub-block for battery electrode material.}
	\label{fig:A2-3}
\end{figure}
\begin{figure}%[htbp]
	\centering
        \includegraphics[scale=.25]{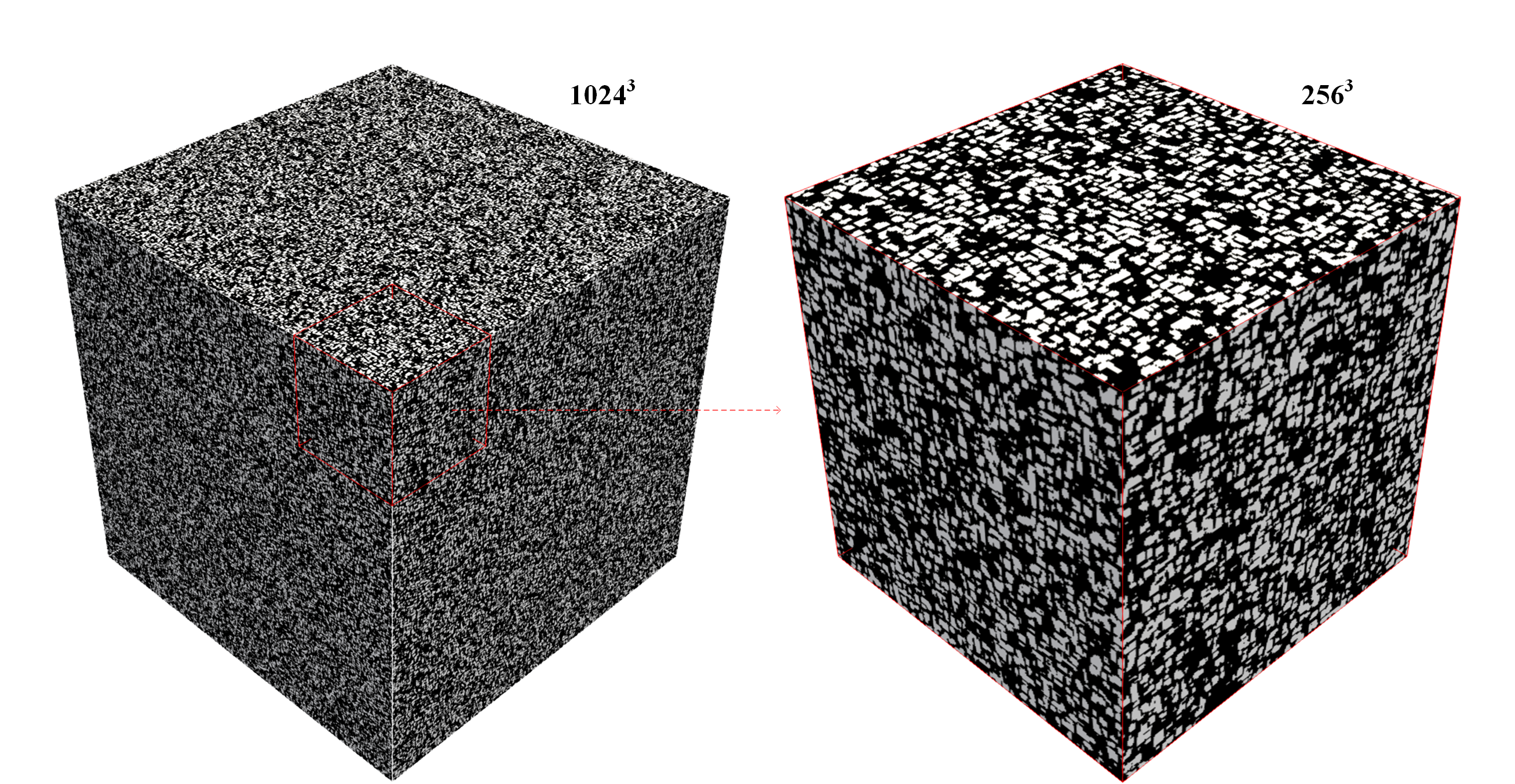}
	\caption{Visualization of a 1024-size reconstruction result and a cropped 256-size sub-block for superalloy.}
	\label{fig:A2-4}
\end{figure}

\newpage
\textbf{A.3 Comparisons of different loss functions}

\begin{figure}%[htbp]
	\centering
        \includegraphics[scale=.36]{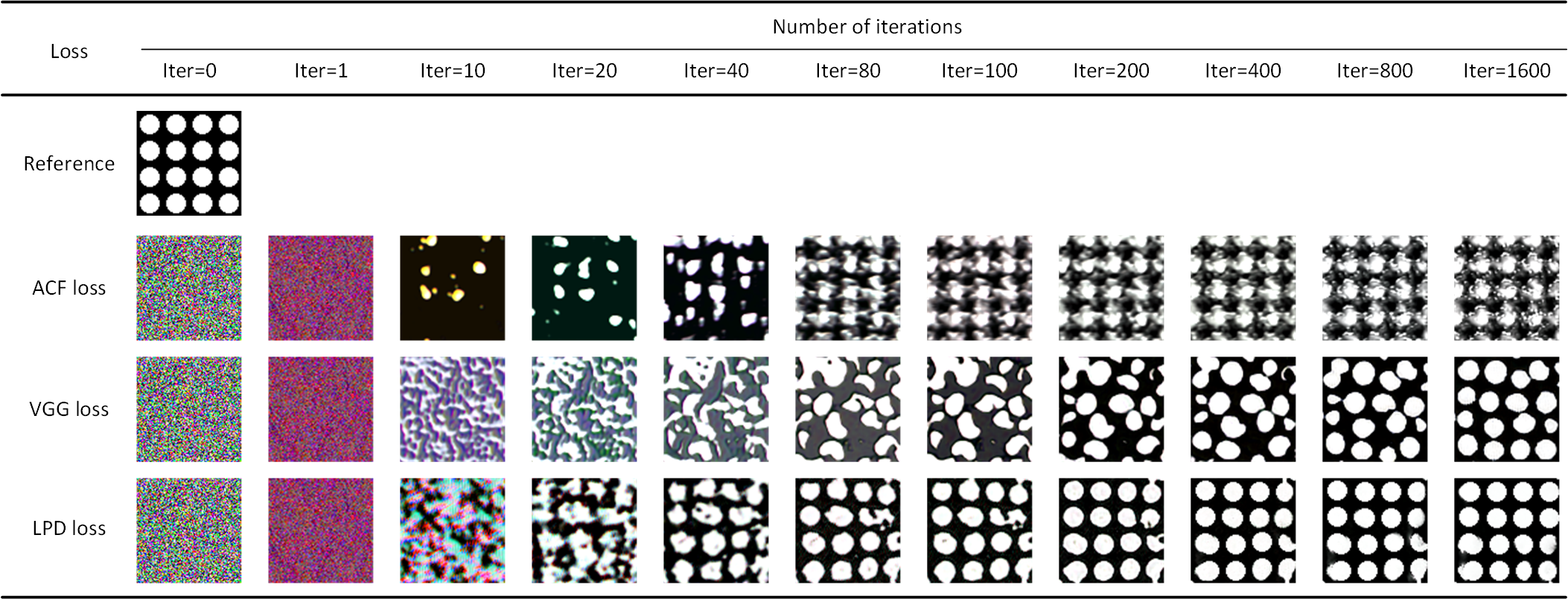}
	\caption{Comparison of the training process among autocorrelation function-based (ACF) loss, VGG network-based (VGG) loss, and local pattern distribution-based (LPD) loss for the regular periodic structure under the same network structure and parameter settings.}
	\label{fig:A3}
\end{figure}

\textbf{A.4 Comparisons of loss iteration processes}

\begin{figure}%[htbp]
	\centering
        \includegraphics[scale=.28]{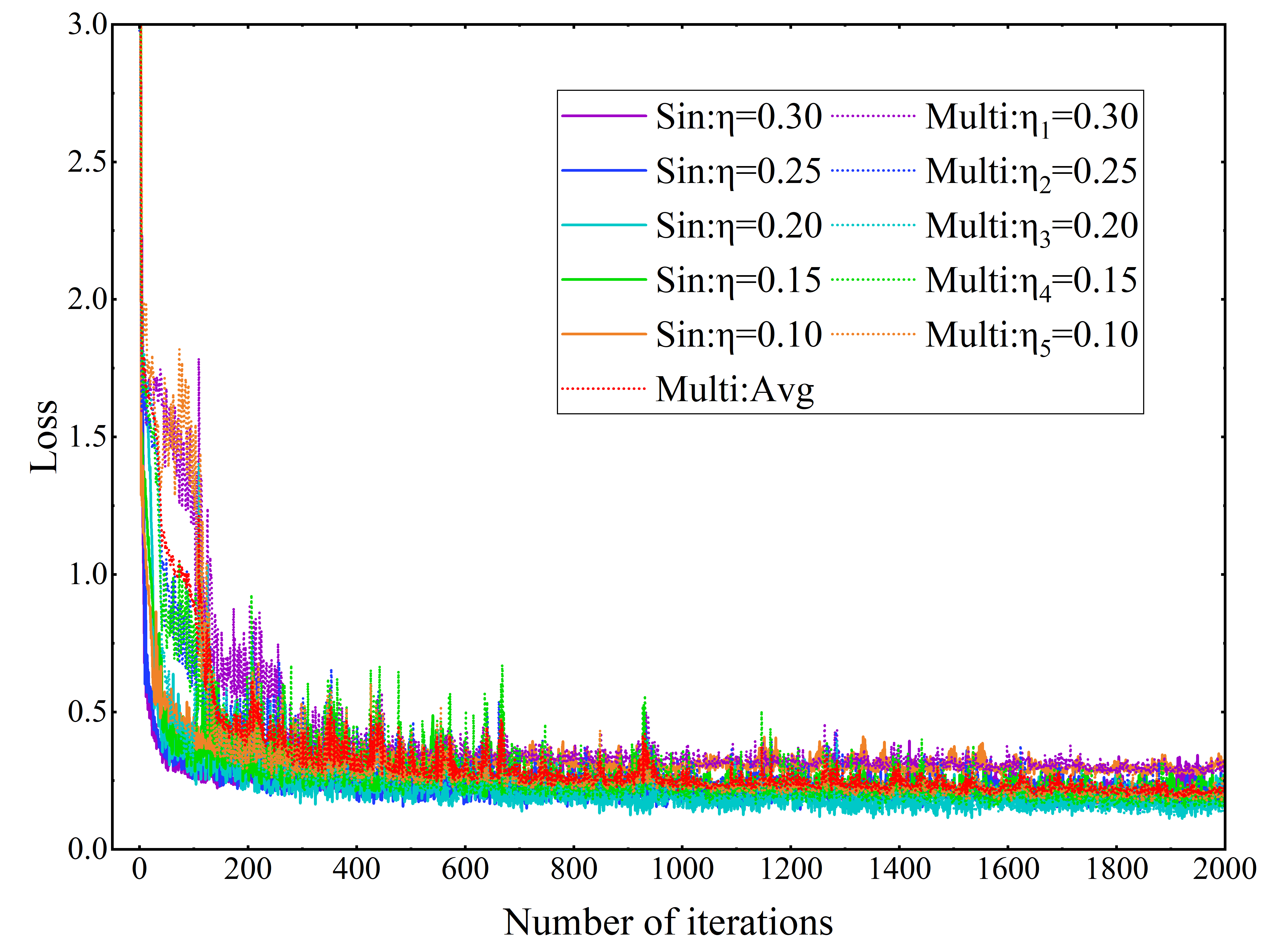}
	\caption{Comparison of loss iteration processes of simultaneously fitting multiple different Gaussian distributions and multiple different target distributions and separately fitting each Gaussian distribution and each target distribution individually. Sin: $\eta =0.30, 0.25, 0.20, 0.15, 0.10$ corresponds to fit single input-to-target mapping separately. Multi: Avg ($\eta _{1-5} =0.30, 0.25, 0.20, 0.15, 0.10$) corresponds to fit multiple input-to-target mappings simultaneously.}
	\label{fig:A4}
\end{figure}

\newpage
\textbf{A.5 Visualization of iteration processes}

\begin{figure}%[htbp]
	\centering
        \includegraphics[scale=.40]{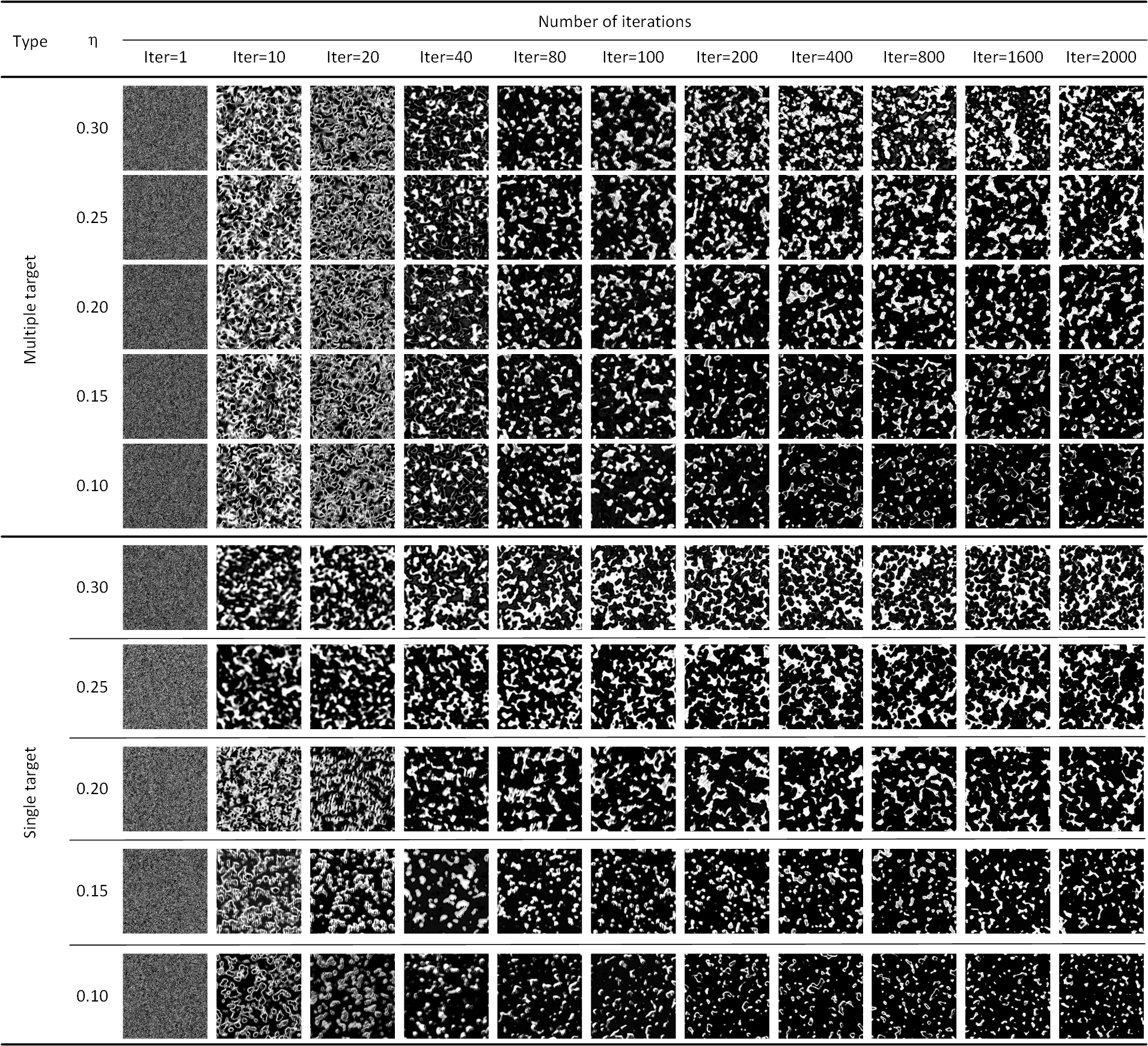}
	\caption{Visualizations of sections in one direction during the iteration process of fitting multiple mappings simultaneously versus fitting a single mapping separately.}
	\label{fig:A5}
\end{figure}

\bibliographystyle{cas-model2-names}
\bibliography{bibliography} 

@article{CBM-hemmati2024acoustic,
  title={Acoustic and thermal performance of wood strands-rock wool-cement composite boards as eco-friendly construction materials},
  author={Hemmati, Negin and Mirzaei, Ramazan and Soltani, Parham and Berardi, Umberto and Mozafari, Mohammad Javad Sheikh and Edalat, Hamidreza and Rezaieyan, Ehsan and Taban, Ebrahim},
  journal={Construction and Building Materials},
  volume={445},
  pages={137935},
  year={2024},
  publisher={Elsevier}
}

@inproceedings{1_farajzadeh2012foam,
  title={Foam-oil interaction in porous media-Implications for foam-assisted enhanced oil recovery (SPE 154197)},
  author={Farajzadeh, Rouhollah and Andrianov, Alexey and Krastev, Rumen and Rossen, WR and Hirasaki, GJ},
  booktitle={74th EAGE Conference and Exhibition incorporating EUROPEC 2012},
  pages={cp--293},
  year={2012},
  organization={European Association of Geoscientists \& Engineers}
}

@article{2_shukla2010review,
  title={A review of studies on CO2 sequestration and caprock integrity},
  author={Shukla, Richa and Ranjith, Pathegama and Haque, Asadul and Choi, Xavier},
  journal={Fuel},
  volume={89},
  number={10},
  pages={2651--2664},
  year={2010},
  publisher={Elsevier}
}

@article{3_hotza2020silicon,
  title={Silicon carbide filters and porous membranes: A review of processing, properties, performance and application},
  author={Hotza, Dachamir and Di Luccio, Marco and Wilhelm, Michaela and Iwamoto, Yuji and Bernard, Samuel and da Costa, Jo{\~a}o C Diniz},
  journal={Journal of Membrane Science},
  volume={610},
  pages={118193},
  year={2020},
  publisher={Elsevier}
}

@article{4_zhang2017melting,
  title={Melting heat transfer characteristics of a composite phase change material fabricated by paraffin and metal foam},
  author={Zhang, Peng and Meng, ZN and Zhu, H and Wang, YL and Peng, SP},
  journal={Applied Energy},
  volume={185},
  pages={1971--1983},
  year={2017},
  publisher={Elsevier}
}

@article{5_lagadec2019characterization,
  title={Characterization and performance evaluation of lithium-ion battery separators},
  author={Lagadec, Marie Francine and Zahn, Raphael and Wood, Vanessa},
  journal={Nature Energy},
  volume={4},
  number={1},
  pages={16--25},
  year={2019},
  publisher={Nature Publishing Group UK London}
}

@article{6_chen2022pore,
  title={Pore-scale modeling of complex transport phenomena in porous media},
  author={Chen, Li and He, An and Zhao, Jianlin and Kang, Qinjun and Li, Zeng-Yao and Carmeliet, Jan and Shikazono, Naoki and Tao, Wen-Quan},
  journal={Progress in Energy and Combustion Science},
  volume={88},
  pages={100968},
  year={2022},
  publisher={Elsevier}
}

@article{7_withers2021x,
  title={X-ray computed tomography},
  author={Withers, Philip J and Bouman, Charles and Carmignato, Simone and Cnudde, Veerle and Grimaldi, David and Hagen, Charlotte K and Maire, Eric and Manley, Marena and Du Plessis, Anton and Stock, Stuart R},
  journal={Nature Reviews Methods Primers},
  volume={1},
  number={1},
  pages={18},
  year={2021},
  publisher={Nature Publishing Group UK London}
}

@article{8_kizilyaprak2014focused,
  title={Focused ion beam scanning electron microscopy in biology},
  author={Kizilyaprak, C and Daraspe, J and Humbel, BM},
  journal={Journal of microscopy},
  volume={254},
  number={3},
  pages={109--114},
  year={2014},
  publisher={Wiley Online Library}
}

@article{9_bostanabad2018computational,
  title={Computational microstructure characterization and reconstruction: Review of the state-of-the-art techniques},
  author={Bostanabad, Ramin and Zhang, Yichi and Li, Xiaolin and Kearney, Tucker and Brinson, L Catherine and Apley, Daniel W and Liu, Wing Kam and Chen, Wei},
  journal={Progress in Materials Science},
  volume={95},
  pages={1--41},
  year={2018},
  publisher={Elsevier}
}

@article{10_zang2024psp,
  title={PSP-GEN: Stochastic inversion of the Process-Structure-Property chain in materials design through deep, generative probabilistic modeling},
  author={Zang, Yaohua and Koutsourelakis, Phaedon-Stelios},
  journal={Acta Materialia},
  pages={120600},
  year={2024},
  publisher={Elsevier}
}

@article{11_shang2023tailoring,
  title={Tailoring the mechanical properties of 3D microstructures: A deep learning and genetic algorithm inverse optimization framework},
  author={Shang, Xiao and Liu, Zhiying and Zhang, Jiahui and Lyu, Tianyi and Zou, Yu},
  journal={Materials Today},
  volume={70},
  pages={71--81},
  year={2023},
  publisher={Elsevier}
}

@article{11_1wang2024diffmat,
  title={DiffMat: Data-driven inverse design of energy-absorbing metamaterials using diffusion model},
  author={Wang, Haoyu and Du, Zongliang and Feng, Fuyong and Kang, Zhong and Tang, Shan and Guo, Xu},
  journal={Computer Methods in Applied Mechanics and Engineering},
  volume={432},
  pages={117440},
  year={2024},
  publisher={Elsevier}
}

@article{JOBE3-kim2022gradient,
  title={Gradient-based phase segmentation method for characterization of hydrating cement paste microstructures obtained from X-ray micro-CT},
  author={Kim, Ji-Su and Suh, Jeewoo and Pae, Junil and Moon, Juhyuk and Han, Tong-Seok},
  journal={Journal of Building Engineering},
  volume={46},
  pages={103721},
  year={2022},
  publisher={Elsevier}
}

@article{12_karsanina2018hierarchical,
  title={Hierarchical optimization: Fast and robust multiscale stochastic reconstructions with rescaled correlation functions},
  author={Karsanina, Marina V and Gerke, Kirill M},
  journal={Physical review letters},
  volume={121},
  number={26},
  pages={265501},
  year={2018},
  publisher={APS}
}

@article{13_zhang2019efficient,
  title={Efficient 3D reconstruction of random heterogeneous media via random process theory and stochastic reconstruction procedure},
  author={Zhang, Wenliang and Song, Lei and Li, Juanjuan},
  journal={Computer Methods in Applied Mechanics and Engineering},
  volume={354},
  pages={1--15},
  year={2019},
  publisher={Elsevier}
}

@article{14_xiao2023novel,
  title={The novel continuous reconstruction approach for reconstructing anisotropic porous rocks},
  author={Xiao, Nan and Zhou, Xiao-Ping},
  journal={Computers and Geotechnics},
  volume={153},
  pages={105101},
  year={2023},
  publisher={Elsevier}
}

@article{15_liu2015random,
  title={Random heterogeneous materials via texture synthesis},
  author={Liu, Xingchen and Shapiro, Vadim},
  journal={Computational Materials Science},
  volume={99},
  pages={177--189},
  year={2015},
  publisher={Elsevier}
}

@article{16_pourfard2017pcto,
  title={PCTO-SIM: Multiple-point geostatistical modeling using parallel conditional texture optimization},
  author={Pourfard, Mohammadreza and Abdollahifard, Mohammad J and Faez, Karim and Motamedi, Sayed Ahmad and Hosseinian, Tahmineh},
  journal={Computers \& Geosciences},
  volume={102},
  pages={116--138},
  year={2017},
  publisher={Elsevier}
}

@article{17_jiang2013efficient,
  title={Efficient 3D porous microstructure reconstruction via Gaussian random field and hybrid optimization},
  author={Jiang, Z and Chen, Wei and Burkhart, C},
  journal={Journal of microscopy},
  volume={252},
  number={2},
  pages={135--148},
  year={2013},
  publisher={Wiley Online Library}
}

@article{18_gao2021ultra,
  title={Ultra-efficient reconstruction of 3D microstructure and distribution of properties of random heterogeneous materials containing multiple phases},
  author={Gao, Yi and Jiao, Yang and Liu, Yongming},
  journal={Acta Materialia},
  volume={204},
  pages={116526},
  year={2021},
  publisher={Elsevier}
}

@article{18_1guo2023spherical,
  title={A spherical harmonic-random field coupled method for efficient reconstruction of CT-image based 3D aggregates with controllable multiscale morphology},
  author={Guo, Fu-qiang and Zhang, Hui and Yang, Zhen-jun and Huang, Yu-jie and Withers, Philip J},
  journal={Computer Methods in Applied Mechanics and Engineering},
  volume={406},
  pages={115901},
  year={2023},
  publisher={Elsevier}
}

@article{19_mosser2017reconstruction,
  title={Reconstruction of three-dimensional porous media using generative adversarial neural networks},
  author={Mosser, Lukas and Dubrule, Olivier and Blunt, Martin J},
  journal={Physical Review E},
  volume={96},
  number={4},
  pages={043309},
  year={2017},
  publisher={APS}
}

@article{20_feng2020end,
  title={An end-to-end three-dimensional reconstruction framework of porous media from a single two-dimensional image based on deep learning},
  author={Feng, Junxi and Teng, Qizhi and Li, Bing and He, Xiaohai and Chen, Honggang and Li, Yang},
  journal={Computer Methods in Applied Mechanics and Engineering},
  volume={368},
  pages={113043},
  year={2020},
  publisher={Elsevier}
}

@article{21_kench2021generating,
  title={Generating three-dimensional structures from a two-dimensional slice with generative adversarial network-based dimensionality expansion},
  author={Kench, Steve and Cooper, Samuel J},
  journal={Nature Machine Intelligence},
  volume={3},
  number={4},
  pages={299--305},
  year={2021},
  publisher={Nature Publishing Group UK London}
}

@article{EWC25-seibert2024fast,
  title={Fast descriptor-based 2D and 3D microstructure reconstruction using the Portilla--Simoncelli algorithm},
  author={Seibert, Paul and Ra{\ss}loff, Alexander and Kalina, Karl and K{\"a}stner, Markus},
  journal={Engineering with Computers},
  pages={1--19},
  year={2024},
  publisher={Springer}
}

@article{21_1kang2024hybrid,
  title={Hybrid LBM and machine learning algorithms for permeability prediction of porous media: A comparative study},
  author={Kang, Qing and Li, Kai-Qi and Fu, Jin-Long and Liu, Yong},
  journal={Computers and Geotechnics},
  volume={168},
  pages={106163},
  year={2024},
  publisher={Elsevier}
}

@article{22_yeong1998reconstructing,
  title={Reconstructing random media},
  author={Yeong, Christofer LY and Torquato, Salvatore},
  journal={Physical review E},
  volume={57},
  number={1},
  pages={495},
  year={1998},
  publisher={APS}
}

@article{23_jiao2014modeling,
  title={Modeling and characterizing anisotropic inclusion orientation in heterogeneous material via directional cluster functions and stochastic microstructure reconstruction},
  author={Jiao, Yang and Chawla, Nikhilesh},
  journal={Journal of Applied Physics},
  volume={115},
  number={9},
  year={2014},
  publisher={AIP Publishing}
}

@article{24_xiao2021texture,
  title={Texture synthesis: A novel method for generating digital models with heterogeneous diversity of rock materials and its CGM verification},
  author={Xiao, Huaiguang and He, Lei and Li, Xing and Zhang, Qianbing and Li, Wengui},
  journal={Computers and Geotechnics},
  volume={130},
  pages={103895},
  year={2021},
  publisher={Elsevier}
}

@article{25_chen20232,
  title={2-D microstructure characterization and reconstruction of heterogeneous materials based on combination of physical descriptor and texture synthesis},
  author={Chen, Yijia and Lin, Li and Sun, Luoming and Xie, Xiyu and Ma, Zhiyuan},
  journal={Materials Characterization},
  volume={196},
  pages={112585},
  year={2023},
  publisher={Elsevier}
}

@article{26_gayon2020pores,
  title={Pores for thought: generative adversarial networks for stochastic reconstruction of 3D multi-phase electrode microstructures with periodic boundaries},
  author={Gayon-Lombardo, Andrea and Mosser, Lukas and Brandon, Nigel P and Cooper, Samuel J},
  journal={npj Computational Materials},
  volume={6},
  number={1},
  pages={82},
  year={2020},
  publisher={Nature Publishing Group UK London}
}

@article{27_zhang2023pm,
  title={PM-ARNN: 2D-TO-3D reconstruction paradigm for microstructure of porous media via adversarial recurrent neural network},
  author={Zhang, Fan and He, Xiaohai and Teng, Qizhi and Wu, Xiaohong and Cui, Junfang and Dong, Xiucheng},
  journal={Knowledge-Based Systems},
  volume={264},
  pages={110333},
  year={2023},
  publisher={Elsevier}
}

@article{28_argilaga2024synthesising,
  title={Synthesising microstructures of a partially frozen salty sand using voxel-based 3D generative adversarial networks},
  author={Argilaga, Albert and Zhao, Chaofa and Li, Hanze and Lei, Liang},
  journal={Computers and Geotechnics},
  volume={170},
  pages={106247},
  year={2024},
  publisher={Elsevier}
}

@article{29_haghverdi2021modified,
  title={A modified simulated annealing algorithm for hybrid statistical reconstruction of heterogeneous microstructures},
  author={Haghverdi, Ali and Baniassadi, Majid and Baghani, Mostafa and Sahraei, Abolfazl Alizadeh and Garmestani, Hamid and Safdari, Masoud},
  journal={Computational Materials Science},
  volume={197},
  pages={110636},
  year={2021},
  publisher={Elsevier}
}

@article{30_yan2023multiscale,
  title={Multiscale reconstruction of porous media based on multiple dictionaries learning},
  author={Yan, Pengcheng and Teng, Qizhi and He, Xiaohai and Ma, Zhenchuan and Zhang, Ningning},
  journal={Computers \& Geosciences},
  volume={176},
  pages={105356},
  year={2023},
  publisher={Elsevier}
}

@article{31_seibert2022descriptor,
  title={Descriptor-based reconstruction of three-dimensional microstructures through gradient-based optimization},
  author={Seibert, Paul and Ra{\ss}loff, Alexander and Ambati, Marreddy and K{\"a}stner, Markus},
  journal={Acta Materialia},
  volume={227},
  pages={117667},
  year={2022},
  publisher={Elsevier}
}

@article{32_ma2024stochastic,
  title={Stochastic reconstruction of heterogeneous microstructure combining sliced Wasserstein distance and gradient optimization},
  author={Ma, Zhenchuan and Teng, Qizhi and Yan, Pengcheng and Wu, Xiaohong and He, Xiaohai},
  journal={Acta Materialia},
  volume={274},
  pages={120023},
  year={2024},
  publisher={Elsevier}
}

@article{32_1fu2021statistical,
  title={Statistical characterization and reconstruction of heterogeneous microstructures using deep neural network},
  author={Fu, Jinlong and Cui, Shaoqing and Cen, Song and Li, Chenfeng},
  journal={Computer Methods in Applied Mechanics and Engineering},
  volume={373},
  pages={113516},
  year={2021},
  publisher={Elsevier}
}

@article{32_2fu2023hierarchical,
  title={Hierarchical reconstruction of 3D well-connected porous media from 2D exemplars using statistics-informed neural network},
  author={Fu, Jinlong and Wang, Min and Xiao, Dunhui and Zhong, Shan and Ge, Xiangyun and Wu, Minglu and Evans, Ben},
  journal={Computer Methods in Applied Mechanics and Engineering},
  volume={410},
  pages={116049},
  year={2023},
  publisher={Elsevier}
}

@article{33_guan2021reconstructing,
  title={Reconstructing porous media using generative flow networks},
  author={Guan, Kelly M and Anderson, Timothy I and Creux, Patrice and Kovscek, Anthony R},
  journal={Computers \& Geosciences},
  volume={156},
  pages={104905},
  year={2021},
  publisher={Elsevier}
}

@article{34_zhang20223d,
  title={3D-PMRNN: Reconstructing three-dimensional porous media from the two-dimensional image with recurrent neural network},
  author={Zhang, Fan and He, Xiaohai and Teng, Qizhi and Wu, Xiaohong and Dong, Xiucheng},
  journal={Journal of Petroleum Science and Engineering},
  volume={208},
  pages={109652},
  year={2022},
  publisher={Elsevier}
}

@article{35_buzzy2024statistically,
  title={Statistically conditioned polycrystal generation using denoising diffusion models},
  author={Buzzy, Michael O and Robertson, Andreas E and Kalidindi, Surya R},
  journal={Acta Materialia},
  volume={267},
  pages={119746},
  year={2024},
  publisher={Elsevier}
}

@article{36_lee2024denoising,
  title={Denoising diffusion-based synthetic generation of three-dimensional (3D) anisotropic microstructures from two-dimensional (2D) micrographs},
  author={Lee, Kang-Hyun and Yun, Gun Jin},
  journal={Computer Methods in Applied Mechanics and Engineering},
  volume={423},
  pages={116876},
  year={2024},
  publisher={Elsevier}
}

@article{37_robertson2023local,
  title={Local--global decompositions for conditional microstructure generation},
  author={Robertson, Andreas E and Kelly, Conlain and Buzzy, Michael and Kalidindi, Surya R},
  journal={Acta Materialia},
  volume={253},
  pages={118966},
  year={2023},
  publisher={Elsevier}
}

@article{38_chi2023reconstruction,
  title={Reconstruction of 3D digital rocks with controllable porosity using CVAE-GAN},
  author={Chi, Peng and Sun, Jianmeng and Luo, Xin and Cui, Ruikang and Dong, Huaimin},
  journal={Geoenergy Science and Engineering},
  volume={230},
  pages={212264},
  year={2023},
  publisher={Elsevier}
}

@article{39_luo2024multi,
  title={A multi-condition denoising diffusion probabilistic model controls the reconstruction of 3D digital rocks},
  author={Luo, Xin and Sun, Jianmeng and Zhang, Ran and Chi, Peng and Cui, Ruikang},
  journal={Computers \& Geosciences},
  volume={184},
  pages={105541},
  year={2024},
  publisher={Elsevier}
}

@article{40_karsanina2023stochastic,
  title={Stochastic (re) constructions of non-stationary material structures: Using ensemble averaged correlation functions and non-uniform phase distributions},
  author={Karsanina, Marina V and Gerke, Kirill M},
  journal={Physica A: Statistical Mechanics and its Applications},
  volume={611},
  pages={128417},
  year={2023},
  publisher={Elsevier}
}

@article{41_xu2022harnessing,
  title={Harnessing structural stochasticity in the computational discovery and design of microstructures},
  author={Xu, Leidong and Hoffman, Nathaniel and Wang, Zihan and Xu, Hongyi},
  journal={Materials \& Design},
  volume={223},
  pages={111223},
  year={2022},
  publisher={Elsevier}
}

@article{42_yang2021exploration,
  title={Exploration of the underlying space in microscopic images via deep learning for additively manufactured piezoceramics},
  author={Yang, Wenhua and Wang, Zhuo and Yang, Tiannan and He, Li and Song, Xuan and Liu, Yucheng and Chen, Lei},
  journal={ACS applied materials \& interfaces},
  volume={13},
  number={45},
  pages={53439--53453},
  year={2021},
  publisher={ACS Publications}
}

@article{43_generale2024inverse,
  title={Inverse stochastic microstructure design},
  author={Generale, Adam P and Robertson, Andreas E and Kelly, Conlain and Kalidindi, Surya R},
  journal={Acta Materialia},
  volume={271},
  pages={119877},
  year={2024},
  publisher={Elsevier}
}

@book{44_li2009markov,
  title={Markov random field modeling in image analysis},
  author={Li, Stan Z},
  year={2009},
  publisher={Springer Science \& Business Media}
}

@article{45_kolouri2017optimal,
  title={Optimal mass transport: Signal processing and machine-learning applications},
  author={Kolouri, Soheil and Park, Se Rim and Thorpe, Matthew and Slepcev, Dejan and Rohde, Gustavo K},
  journal={IEEE signal processing magazine},
  volume={34},
  number={4},
  pages={43--59},
  year={2017},
  publisher={IEEE}
}

@book{46_kroner1972statistical,
  title={Statistical continuum mechanics},
  author={Kr{\"o}ner, Ekkehart},
  volume={92},
  year={1972},
  publisher={Springer}
}

@article{47_torquato1997effective,
  title={Effective stiffness tensor of composite media—I. Exact series expansions},
  author={Torquato, Salvatore},
  journal={Journal of the Mechanics and Physics of Solids},
  volume={45},
  number={9},
  pages={1421--1448},
  year={1997},
  publisher={Elsevier}
}

@article{48_torquato1998effective,
  title={Effective stiffness tensor of composite media: II. Applications to isotropic dispersions},
  author={Torquato, Salvatore},
  journal={Journal of the Mechanics and Physics of Solids},
  volume={46},
  number={8},
  pages={1411--1440},
  year={1998},
  publisher={Elsevier}
}

@article{49_torquato2002statistical,
  title={Statistical description of microstructures},
  author={Torquato, Salvatore},
  journal={Annual review of materials research},
  volume={32},
  number={1},
  pages={77--111},
  year={2002},
  publisher={Annual Reviews 4139 El Camino Way, PO Box 10139, Palo Alto, CA 94303-0139, USA}
}

@article{50_cheng2022data,
  title={Data-driven learning of 3-point correlation functions as microstructure representations},
  author={Cheng, Sheng and Jiao, Yang and Ren, Yi},
  journal={Acta Materialia},
  volume={229},
  pages={117800},
  year={2022},
  publisher={Elsevier}
}

@article{51_postnicov2024evaluation,
  title={Evaluation of three-point correlation functions from structural images on CPU and GPU architectures: Accounting for anisotropy effects},
  author={Postnicov, Vasily and Karsanina, Marina V and Khlyupin, Aleksey and Gerke, Kirill M},
  journal={Physical Review E},
  volume={110},
  number={4},
  pages={045306},
  year={2024},
  publisher={APS}
}

@article{52_li2006quantitative,
  title={Quantitative prediction of effective conductivity in anisotropic heterogeneous media using two-point correlation functions},
  author={Li, DS and Saheli, Ghazal and Khaleel, M and Garmestani, Hamid},
  journal={Computational Materials Science},
  volume={38},
  number={1},
  pages={45--50},
  year={2006},
  publisher={Elsevier}
}

@article{53_belvin2009application,
  title={Application of two-point probability distribution functions to predict properties of heterogeneous two-phase materials},
  author={Belvin, A and Burrell, R and Gokhale, A and Thadhani, N and Garmestani, H},
  journal={Materials characterization},
  volume={60},
  number={9},
  pages={1055--1062},
  year={2009},
  publisher={Elsevier}
}

@article{54_roding2017functional,
  title={Functional regression-based fluid permeability prediction in monodisperse sphere packings from isotropic two-point correlation functions},
  author={R{\"o}ding, Magnus and Svensson, Peter and Lor{\'e}n, Niklas},
  journal={Computational materials science},
  volume={134},
  pages={126--131},
  year={2017},
  publisher={Elsevier}
}

@article{55_ding2018improved,
  title={Improved multipoint statistics method for reconstructing three-dimensional porous media from a two-dimensional image via porosity matching},
  author={Ding, Kai and Teng, Qizhi and Wang, Zhengyong and He, Xiaohai and Feng, Junxi},
  journal={Physical Review E},
  volume={97},
  number={6},
  pages={063304},
  year={2018},
  publisher={APS}
}

@article{56_xia2021three,
  title={Three-dimensional reconstruction of porous media using super-dimension-based adjacent block-matching algorithm},
  author={Xia, Zhixin and Teng, Qizhi and Wu, Xiaohong and Li, Juan and Yan, Pengcheng},
  journal={Physical Review E},
  volume={104},
  number={4},
  pages={045308},
  year={2021},
  publisher={APS}
}

@article{57_ma2023fast,
  title={A fast and flexible algorithm for microstructure reconstruction combining simulated annealing and deep learning},
  author={Ma, Zhenchuan and He, Xiaohai and Yan, Pengcheng and Zhang, Fan and Teng, Qizhi},
  journal={Computers and Geotechnics},
  volume={164},
  pages={105755},
  year={2023},
  publisher={Elsevier}
}

@inproceedings{58_DBLP:journals/corr/KingmaB14,
  author       = {Diederik P. Kingma and
                  Jimmy Ba},
  title        = {Adam: {A} Method for Stochastic Optimization},
  booktitle    = {3rd International Conference on Learning Representations},
  year         = {2015},
}

@article{59_bostanabad2016characterization,
  title={Characterization and reconstruction of 3D stochastic microstructures via supervised learning},
  author={Bostanabad, Ramin and Chen, Wei and Apley, Daniel W},
  journal={Journal of microscopy},
  volume={264},
  number={3},
  pages={282--297},
  year={2016},
  publisher={Wiley Online Library}
}

@article{60_coker1996morphology,
  title={Morphology and physical properties of Fontainebleau sandstone via a tomographic analysis},
  author={Coker, David A and Torquato, Salvatore and Dunsmuir, John H},
  journal={Journal of Geophysical Research: Solid Earth},
  volume={101},
  number={B8},
  pages={17497--17506},
  year={1996},
  publisher={Wiley Online Library}
}

@article{61_feng2018reconstruction,
  title={Reconstruction of three-dimensional heterogeneous media from a single two-dimensional section via co-occurrence correlation function},
  author={Feng, Junxi and Teng, Qizhi and He, Xiaohai and Qing, Linbo and Li, Yang},
  journal={Computational Materials Science},
  volume={144},
  pages={181--192},
  year={2018},
  publisher={Elsevier}
}

@article{62_ferguson2018puma,
  title={PuMA: The porous microstructure analysis software},
  author={Ferguson, Joseph C and Panerai, Francesco and Borner, Arnaud and Mansour, Nagi N},
  journal={SoftwareX},
  volume={7},
  pages={81--87},
  year={2018},
  publisher={Elsevier}
}

@article{63_ferguson2021update,
  title={Update 3.0 to “PuMA: The porous microstructure analysis software”,(PII: S2352711018300281)},
  author={Ferguson, Joseph C and Semeraro, Federico and Thornton, John M and Panerai, Francesco and Borner, Arnaud and Mansour, Nagi N},
  journal={SoftwareX},
  volume={15},
  pages={100775},
  year={2021},
  publisher={Elsevier}
}

@book{64_hestenes1952methods,
  title={Methods of conjugate gradients for solving linear systems},
  author={Hestenes, Magnus Rudolph and Stiefel, Eduard and others},
  volume={49},
  number={1},
  year={1952},
  publisher={NBS Washington, DC}
}

@article{65_mcnamara1988use,
  title={Use of the Boltzmann equation to simulate lattice-gas automata},
  author={McNamara, Guy R and Zanetti, Gianluigi},
  journal={Physical review letters},
  volume={61},
  number={20},
  pages={2332},
  year={1988},
  publisher={APS}
}

@article{66_santos2022mplbm,
  title={MPLBM-UT: Multiphase LBM library for permeable media analysis},
  author={Santos, Javier E and Gigliotti, Alex and Bihani, Abhishek and Landry, Christopher and Hesse, Marc A and Pyrcz, Michael J and Prodanovi{\'c}, Ma{\v{s}}a},
  journal={SoftwareX},
  volume={18},
  pages={101097},
  year={2022},
  publisher={Elsevier}
}

@article{67_latt2021palabos,
  title={Palabos: parallel lattice Boltzmann solver},
  author={Latt, Jonas and Malaspinas, Orestis and Kontaxakis, Dimitrios and Parmigiani, Andrea and Lagrava, Daniel and Brogi, Federico and Belgacem, Mohamed Ben and Thorimbert, Yann and Leclaire, S{\'e}bastien and Li, Sha and others},
  journal={Computers \& Mathematics with Applications},
  volume={81},
  pages={334--350},
  year={2021},
  publisher={Elsevier}
}

@article{68_murgas2024modeling,
  title={Modeling complex polycrystalline alloys using a Generative Adversarial Network enabled computational platform},
  author={Murgas, Brayan and Stickel, Joshua and Brewer, Luke and Ghosh, Somnath},
  journal={Nature Communications},
  volume={15},
  number={1},
  pages={9441},
  year={2024},
  publisher={Nature Publishing Group UK London}
}

@inproceedings{69_DBLP:journals/corr/SimonyanZ14a,
  author       = {Karen Simonyan and
                  Andrew Zisserman},
  title        = {Very Deep Convolutional Networks for Large-Scale Image Recognition},
  booktitle    = {3rd International Conference on Learning Representations},
  year         = {2015},
}

@article{70_millan2022study,
  title={Study of a Bimodal $\alpha$--$\beta$ Ti Alloy Microstructure Using Multi-Resolution Spherical Indentation Stress-Strain Protocols},
  author={Millan-Espitia, Natalia and Kalidindi, Surya R},
  journal={Journal of Composites Science},
  volume={6},
  number={6},
  pages={162},
  year={2022},
  publisher={MDPI}
}

\end{document}